\documentclass[aps,pra,amsmath,amssymb,superscriptaddress, reprint]{revtex4-2}
\usepackage{xr}
\usepackage{graphicx}
\usepackage{fixme}
\fxsetup{status=draft}
\usepackage{dcolumn}
\usepackage{bm}
\usepackage{lipsum}
\usepackage{todonotes}
\usepackage{siunitx}
\usepackage{upgreek}
\usepackage{fixme}
\usepackage{todonotes}
\newcommand{\nm}{\mathrm{nm}}
\newcommand{\um}{\mathrm{\mu m}}

\newcolumntype{L}{>{\centering\arraybackslash}p{0.2\linewidth}}
\newcolumntype{M}{>{\centering\arraybackslash}p{0.7\linewidth}}

\begin{document}

\title[practical differential single pixel]{Impedance-matched differential superconducting nanowire detectors}
%\thanks{Footnote to title of article.}

\author{Marco Colangelo}
 \email{colang@mit.edu}
 \affiliation{Department of Electrical Engineering and Computer Science,
Massachusetts Institute of Technology, Cambridge, MA, USA.}%

\author{Boris Korzh}
\email{bkorzh@jpl.caltech.edu}
\affiliation{Jet Propulsion Laboratory, California Institute of Technology, 4800 Oak Grove Dr., Pasadena, CA, USA}

\author{Jason P. Allmaras}
\affiliation{Jet Propulsion Laboratory, California Institute of Technology, 4800 Oak Grove Dr., Pasadena, CA, USA}
\affiliation{Applied Physics, California Institute of Technology, 1200 E California Blvd, Pasadena, CA, USA}

\author{Andrew D. Beyer}%
\affiliation{Jet Propulsion Laboratory, California Institute of Technology, 4800 Oak Grove Dr., Pasadena, CA, USA}

\author{Andrew S. Mueller}
\affiliation{Jet Propulsion Laboratory, California Institute of Technology, 4800 Oak Grove Dr., Pasadena, CA, USA}
\affiliation{Applied Physics, California Institute of Technology, 1200 E California Blvd, Pasadena, CA, USA}

\author{Ryan M. Briggs}
\affiliation{Jet Propulsion Laboratory, California Institute of Technology, 4800 Oak Grove Dr., Pasadena, CA, USA}
 
\author{Bruce Bumble}
\affiliation{Jet Propulsion Laboratory, California Institute of Technology, 4800 Oak Grove Dr., Pasadena, CA, USA}

\author{Marcus Runyan}
\affiliation{Jet Propulsion Laboratory, California Institute of Technology, 4800 Oak Grove Dr., Pasadena, CA, USA}

\author{Martin J. Stevens}
\affiliation{National Institute of Standards and Technology, Boulder, CO, USA.}

\author{Adam N. McCaughan}
\affiliation{National Institute of Standards and Technology, Boulder, CO, USA.}

\author{Di Zhu}
 \affiliation{Department of Electrical Engineering and Computer Science,
Massachusetts Institute of Technology, Cambridge, MA, USA.}%

\author{Stephen Smith}
\affiliation{Cosmic Microwave Technology, 15703 Condon Avenue, Lawndale, CA, USA
}%

\author{Wolfgang Becker}
\affiliation{Becker \& Hickl GmbH, Nahmitzer Damm 30, Berlin, Germany}

\author{Lautaro Narv\'aez}
\affiliation{Division of Physics, Mathematics and Astronomy, California Institute of Technology, Pasadena, CA, USA}%

\author{Joshua C. Bienfang}
\affiliation{National Institute of Standards and Technology, Gaithersburg, MD, USA.}%

\author{Simone Frasca}
\affiliation{Jet Propulsion Laboratory, California Institute of Technology, 4800 Oak Grove Dr., Pasadena, CA, USA}

\author{Angel E. Velasco}
\affiliation{Jet Propulsion Laboratory, California Institute of Technology, 4800 Oak Grove Dr., Pasadena, CA, USA}

\author{Cristi\'an H. Pe\~na}
\affiliation{Fermi National Accelerator Laboratory, Batavia, IL, USA}%
\affiliation{Division of Physics, Mathematics and Astronomy, California Institute of Technology, Pasadena, CA, USA}%

\author{Edward E. Ramirez}
\affiliation{Jet Propulsion Laboratory, California Institute of Technology, 4800 Oak Grove Dr., Pasadena, CA, USA}
\affiliation{California State University, Los Angeles, CA, USA}

\author{Alexander B. Walter}
\affiliation{Jet Propulsion Laboratory, California Institute of Technology, 4800 Oak Grove Dr., Pasadena, CA, USA}

\author{Ekkehart Schmidt}
\affiliation{Jet Propulsion Laboratory, California Institute of Technology, 4800 Oak Grove Dr., Pasadena, CA, USA}

\author{Emma E. Wollman}
\affiliation{Jet Propulsion Laboratory, California Institute of Technology, 4800 Oak Grove Dr., Pasadena, CA, USA}

\author{Maria Spiropulu}
\affiliation{Division of Physics, Mathematics and Astronomy, California Institute of Technology, Pasadena, CA, USA}%

\author{Richard Mirin}
\affiliation{National Institute of Standards and Technology, Boulder, CO, USA.}%

\author{Sae Woo Nam}
\affiliation{National Institute of Standards and Technology, Boulder, CO, USA.}%

\author{Karl K. Berggren}
 \affiliation{Department of Electrical Engineering and Computer Science,
Massachusetts Institute of Technology, Cambridge, MA, USA.}%

\author{Matthew D. Shaw}
\affiliation{Jet Propulsion Laboratory, California Institute of Technology, 4800 Oak Grove Dr., Pasadena, CA, USA}

\date{\today}% It is always \today, today,

\begin{abstract}
Superconducting nanowire single-photon detectors (SNSPDs) are the highest performing photon-counting technology in the near-infrared (NIR). Due to delay-line effects, large area SNSPDs typically trade-off timing resolution and detection efficiency. Here, we introduce a detector design based on transmission line engineering and differential readout for device-level signal conditioning, enabling  a high system detection efficiency and a low detector jitter, simultaneously. To make our differential detectors compatible with single-ended time taggers, we also engineer analog differential-to-single-ended readout electronics, with minimal impact on the system timing resolution. Our niobium nitride differential SNSPDs achieve $47.3\,\% \pm 2.4\,\%$ system detection efficiency and sub-10\,\si{ps} system jitter at $775\,\si{nm}$, while at $1550\,\si{nm}$ they achieve $71.1\,\% \pm 3.7\,\%$ system detection efficiency and $13.1\,\si{ps} \pm 0.4\,\si{ps}$ system jitter. These detectors also achieve sub-100 ps timing response at one one-hundredth maximum level, $30.7\,\si{ps} \pm 0.4\,\si{ps}$ at $775\,\si{nm}$ and $47.6\,\si{ps} \pm 0.4\,\si{ps}$ at $1550\,\si{nm}$, enabling time-correlated single-photon counting with high dynamic range response functions. Furthermore, thanks to the differential impedance-matched design, our detectors exhibit delay-line imaging capabilities and photon-number resolution. The properties and high-performance metrics achieved by our system make it a versatile photon-detection solution for many scientific applications.  
\end{abstract}

\maketitle

\section{Introduction}

\begin{figure*}
	\centering
	\includegraphics{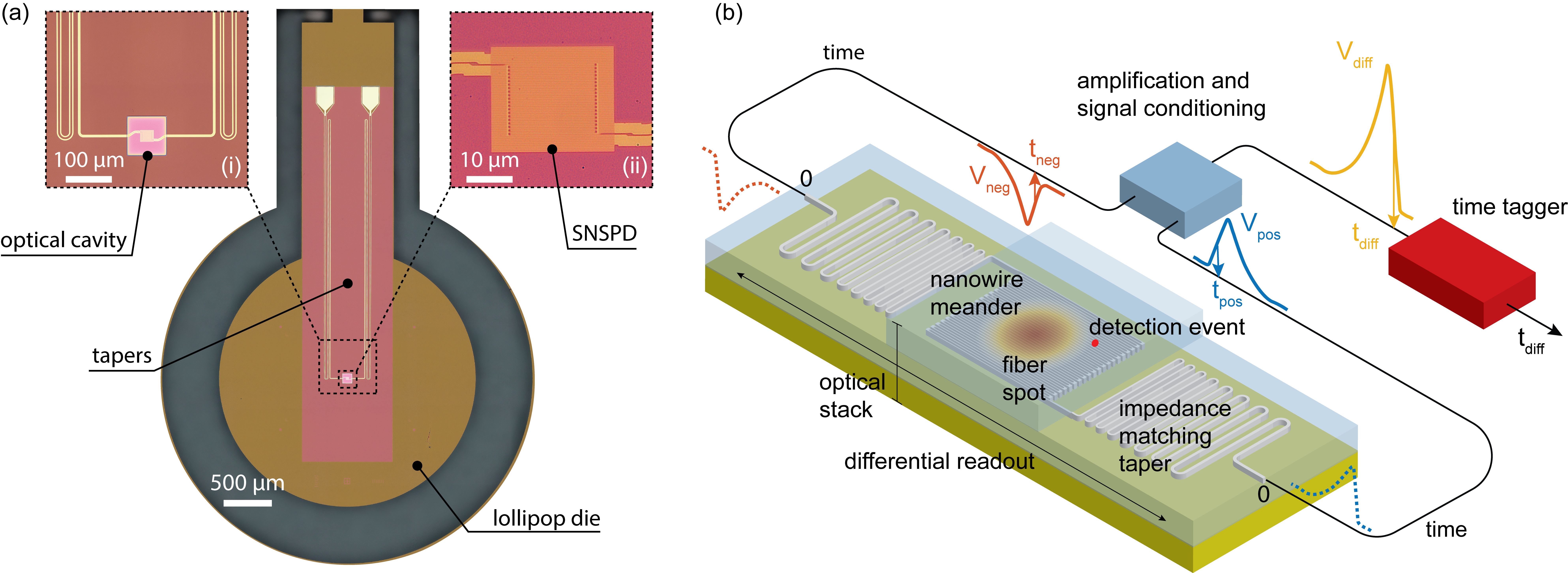}
	\caption{Differential single-pixel superconducting nanowire single-photon detector. (a) Optical micrograph of a representative device. The detector die (lollipop) connected to the parent wafer before packaging. Inset (i): Optical micrograph of the detector. The nanowire is embedded in an optical cavity to improve the detection efficiency. Inset (ii): Optical micrograph of the detector active area. (b) Sketch of the SNSPD architecture realized in this work. Each meander end is interfaced to the readout though an impedance matching taper. Note that the proportions of the several elements of the figure are not to scale. Each detection event generates two complementary rising edges $V_\mathrm{pos}$ and $V_\mathrm{neg}$. The dashed lines show the evolution of the detection pulses in the impedance-matching tapers, before arrival at the amplification and signal conditioning electronics. The pulses are processed to produce their difference $V_\mathrm{diff}$, which is then fed to a single-ended TCSPC module. The TCSPC module returns the time tag $t_\mathrm{diff}$, which has the geometric contribution to the timing jitter compensated for.}
	\label{fig:architecture}
\end{figure*}

Superconducting nanowire single photon detectors (SNSPD) are the preferred photon-counting technology in the near-infrared (NIR). Specialized SNSPD designs can achieve $98\,\%$ efficiency at telecom wavelengths\,\cite{reddy2020superconducting}, ultra-low intrinsic dark count rates\,\cite{wollman2017uv, hochberg2019detecting}, single element count rates $>$ 100\,MHz\,\cite{kerman2013readout}, intrinsic timing jitter as low as 2.6\,\si{ps} to 4.3\,\si{ps}, depending on the wavelength\,\cite{korzh2020demonstration} as well as intrinsic photon-number resolution\,\cite{zhu2020resolving}. Combining several of these metrics into a single device is desirable in many quantum communication applications\,\cite{hadfield2009single} such as long distance\,\cite{boaron2018secure, shibata2014quantum} and high-clock-rate quantum key distribution\,\cite{takesue2007quantum, grunenfelder2020performance}, quantum teleportation\,\cite{valivarthi2020teleportation, chapman2020time-bin}, entanglement swapping\,\cite{samara2020entanglement} and multiplexed single photon generation\,\cite{Kaneda2019high-efficiency}. Detectors excelling in several metrics can also have huge impact in other fields, such as laser ranging\,\cite{mccarthy2013kilometre}, low-power optical waveform capture\,\cite{wang2019oscilloscopic}, fluorescence lifetime imaging\,\cite{becker2005time-correlated}, time-domain diffuse correlation-spectroscopy\,\cite{sutin2016time} and deep-space optical communication\,\cite{shaw2017superconducting}.

One of the main challenges is the combination of an active area large enough for efficient coupling to a single-mode fiber together with low-jitter operation. In fact, the pulse propagation delays induced by the low phase velocity in long nanowires can contribute several tens of ps to the uncertainty of the timing of the readout signal \,\cite{zhao2017single, calandri2016superconducting}. Therefore, large-area detectors with high detection efficiencies do not typically exhibit few-ps timing resolution.  

Another challenge stands in the way of the introduction of low-jitter SNSPDs into applications. Although a few detector designs achieve sub-10 ps timing resolution, these investigations typically make use of high-bandwidth oscilloscopes to characterize only the detector jitter\,\cite{korzh2020demonstration, esmaeilzadeh2020efficient}. In real applications, detectors are deployed with readout electronics which tend to degrade the overall system jitter. Therefore, to achieve a high-system timing resolution, the introduction of high-performance time tagging\,\cite{shcheslavskiy2016ultrafast} and signal conditioning is also critical.

To address the combination of large-area with low-jitter operation, in this paper we engineer impedance-matched devices in a differential configuration. Our device inherits its base elements from the superconducting nanowire single-photon imager (SNSPI)\,\cite{zhao2017single}. The SNSPI was originally designed to provide micrometer-level spatial resolution based on the timing information of photon-detection pulses. The impedance matching tapers were used to preserve the fast-rising edges of the pulses, and the detection events' locations were extracted through post-processing based on the differential readout. Here, we engineer these elements to significantly improve the system performances of single-pixel SNSPDs over traditional designs. By single-pixel SNSPDs, we refer to nanowire detectors with active areas larger than $10\,\mathrm{\mu m} \times 10\,\mathrm{\mu m}$, and embedded in optical stacks tuned to maximize the system detection efficiency. Inspired by our previous proof-of-concept impedance-matched SNSPD\,\cite{zhu2019superconducting}, our impedance-matching tapers are specifically designed to achieve a superior signal-to-noise ratio, and minimize reflections and distortions at the device level. The differential readout architecture, previously shown with lumped devices \,\cite{calandri2016superconducting}, is here applied to impedance matched devices and it is engineered to cancel the geometric delay-line contributions to the timing jitter, achieving a lower detector jitter. This optimized architecture provides a path to low-jitter large-area single-pixel designs, breaking the existing trade-off between these design variables.

The impedance-matched differential architecture of this device offers additional advantages. By operating this single-pixel SNSPD as an SNSPI\,\cite{zhao2017single}, e.g. using the differential readout to determine the coordinate of the detection event, we achieve delay-line imaging capabilities. Moreover, our impedance-matched differential detectors achieve photon-number resolution capabilities up to three photons, following our previous proof-of-concept with a single-ended device\, \cite{zhu2020resolving}. 

For optimal operation, this detector architecture would require two high-performance low-jitter time taggers, which is impractical at scale; to overcome this limitation, we designed a differential-to-single-ended readout system based on analog electronics, which automatically cancels the geometric contribution and minimizes the overall system jitter. The pulses are first amplified and conditioned through analog electronics. The processed pulse is then fed to a time-correlated single-photon counting (TCSPC) module that completes the tagging operation and outputs a time-tag with the delay-line contribution compensated for. Our readout scheme makes the differential detector compatible with traditional single-ended readout system while maintaining the advantages of the optimized architecture.

In Figure \ref{fig:architecture}(a) we show an optical micrograph of one of our impedance-matched differential detector before packaging. The detector is fiber-coupled through a self-aligned packaging method using lollipop-shaped dies \cite{miller2011compact}. The superconducting tapers, interfacing the nanowire to the $50\,\mathrm{\Omega}$ readout, extend along the die. The superconducting nanowire is arranged as a meander and it is embedded in an optical cavity with a gold reflector to maximize photon absorption (inset (i) and (ii)). See Appendix \ref{app_microwave} for details of the microwave design and Appendix\,\ref{fab} for details of the fabrication. Figure \ref{fig:architecture}(b) shows a sketch of the overall architecture of our detection system. The impedance-matched nanowire meander is laid out in a differential configuration. A detection event generates  a positive and a negative pulse, $V_\mathrm{pos}$ and $V_\mathrm{neg}$ respectively. The pulses are amplified and processed to produce their difference, $V_\mathrm{diff}$, which is then fed to a single-ended TCSPC module.

By combining our impedance-matched differential SNSPD with cryogenic differential-to-single-ended readout electronics and a time-to-analog-based time tagger, we achieved system jitter as low as $7.3\,\si{ps} \pm 0.3\,\si{ps}$ full-width at half maximum (FWHM) for straight nanowires. For nanowire devices with active areas larger than $15\,\si{\mu m}\times 10\,\si{\mu m}$, we achieved system detection efficiency higher than $45\,\%$ at both 775\,nm and 1550\,nm. The FWHM system jitter were $9.7\,\si{ps} \pm 0.4\,\si{ps}$  and $13.1\,\si{ps} \pm 0.4\,\si{ps}$, respectively. With the same measurement setup we also achieved a state-of-the-art timing response at the 1/100-of-maximum level (FW1/100M).      

The delay-line imaging capabilities of our detectors are particularly useful to identify the nature of the source of illumination conditions, to debug fiber coupling alignment, assess fabrication yield, estimate upper bounds for detection efficiency, and determine the effective phase velocity of the detector. The photon number resolution capabilities enable further applications making this detector extremely versatile for several measurement scenarios.

The paper is organized as follows. In Sec.\,\ref{sec_SNSPD_design} we discuss our design approach. We show how our optimized architecture can break the trade-off between high system detection efficiency and low detector jitter in traditional SNSPD designs (Sec.\,\ref{sec_SNSPD_design}(A)). We also discuss readout electronics and measurement setups that preserve the detector timing resolution at the system level (Sec.\,\ref{sec_SNSPD_design}(B)). The implementation of the design and the characterization of detectors performance are presented in Sec.\,\ref{sec_result}. Here, we show the detector output pulse and we discuss, compare, and validate the time tagging procedure (Sec.\,\ref{sec_result}(A)). We then present the metrics achieved for the system detection efficiency and we correlate the results to the alignment of the fiber spot on the active area exploiting the imaging capabilities of the detector (Sec.\,\ref{sec_result}(B)). We then discuss the detector and system timing resolution (Sec.\,\ref{sec_result}(C)), and we show how our detector achieves photon-number resolution (Sec.\,\ref{sec_result}(D)). A general discussion on the results can be found in Sec.\,\ref{sec_discussion}. Finally, in Sec.\,\ref{sec_outlook}, we present a summary of the work and discuss applications and impact of this detector on several fields of science.

\section{Approach}\label{sec_SNSPD_design} 

In the following, we propose a new design based on the microwave engineering of superconducting nanowires, enabling high system detection efficiency and low detector jitter, at the same time (Sec.\,\ref{sec_SNSPD_design}(A)), overcoming the limitation of traditional devices. We then discuss how specialized readout electronics can be used to preserve timing resolution at the system level (Sec.\,\ref{sec_SNSPD_design}(B)). To facilitate the reader's understanding of the content, in Table \ref{tab:formulas} we provide an overview of the terminology adopted in this and next sections.  
\begin{table}
	\centering
	\begin{tabular}{L|M}
		Parameter & Description \\ \hline \hline
		$t_\mathrm{pos}$ & time tag of the positive pulse $V_\mathrm{pos}$  \\ \hline
		$t_\mathrm{neg}$ & time tag of the negative pulse $V_\mathrm{neg}$ \\ \hline		
		$t_\mathrm{\Sigma}$ & sum of the time tags $t_\mathrm{\Sigma} = \frac{t_\mathrm{pos}+t_\mathrm{neg}}{2}$ \\ \hline
		$t_\Delta$ & difference of the time tags $t_\mathrm{\Delta} = t_\mathrm{pos}-t_\mathrm{neg}$ \\  \hline
		$\Delta t_\Delta$ & difference between adjacent sub-peaks in the $t_\Delta$ histogram \\ \hline
		$t_\mathrm{x}$ & difference in consecutive $\Delta t_\Delta$, encoding  horizontal shift from the center of the detector ($\Delta x$) \\ \hline
		$t_\mathrm{y}$ & $t_\Delta$ histogram peak, encoding vertical shift from the center of the detector ($\Delta y$) \\ \hline
		$t_\mathrm{diff}$ & time tag of the difference of the complementary pulses $V_\mathrm{diff}=V_\mathrm{pos}-V_\mathrm{neg}$  \\  \hline
		$j_\mathrm{geom}$ & geometric contribution to the timing jitter \\  \hline
		$j_\mathrm{amp}$ & readout electrical noise contribution to the timing jitter \\  \hline
		$j_\Sigma$ & detector jitter, associated to $t_\mathrm{\Sigma}$ \\  \hline
		$j_\mathrm{diff}$ & system jitter, associated to $t_\mathrm{diff}$ \\  \hline

	\end{tabular}
	\caption{Summary of relevant timing parameters and definitions.}
	\label{tab:formulas}
\end{table}

\subsection{Device design: impedance-matched differential SNSPD}
\label{sec_dev_sim}

Single-pixel SNSPDs are traditionally modeled as lumped elements where the nanowire, usually a meander covering tens of \si{\upmu m^2}, behaves as a photon-triggered time-dependent resistor in series with a kinetic inductor\,\cite{berggren2018superconducting}. Although providing a first-level description of the electrothermal dynamics, this picture does not consider the pulse propagation properties in the meander \cite{santavicca2016microwave}. To achieve an optimal design, microwave properties need to be accounted for. 
\begin{figure*}[!ht]
	\centering
	\includegraphics{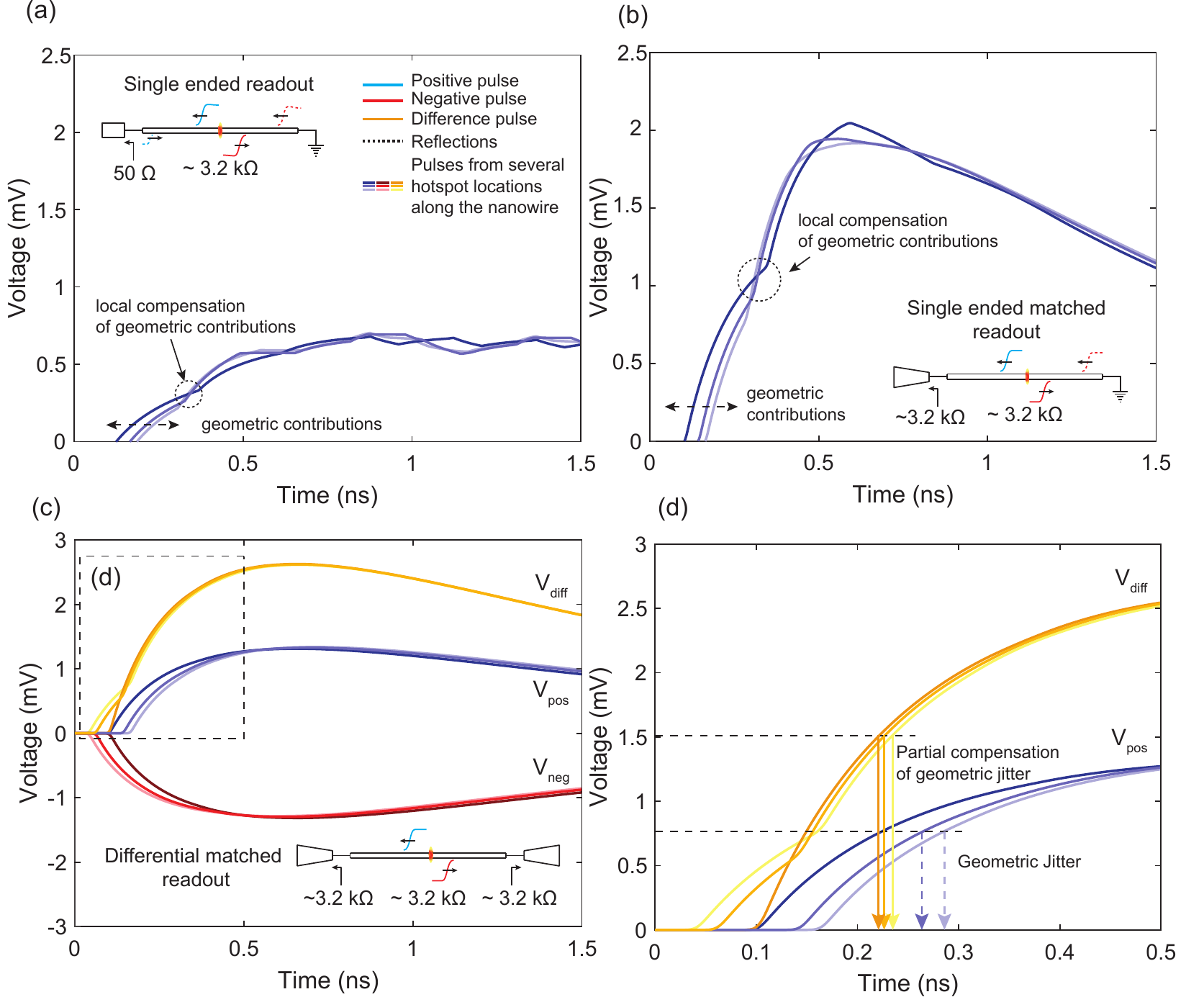}
	\caption{Simulated SNSPD dynamics in several readout configuration. For each sub-figure, the color of the curves (grayscale dark to light) indicates the pulses generated by photon absorption in different locations of the meander (near the center to near the edge). The schematics shown in the insets of sub-figures (a-c) show the simulated readout setups. For simplicity, the nanowire meander is pictured as a straight wire. The blue (red) lines indicate the positive (negative) pulses generated by the photon absorption in the wire. The dashed lines indicate the reflection of the original pulses generated by impedance mismatch. (a) The SNSPD is configured for single-ended readout. The impedance mismatch induced by the 50 \si{\Omega} readout and ground termination generate reflections and distortions which affect the signal-to-noise ratio (SNR). Moreover, due to the slow phase velocity, the relative delay induced by detection in different locations of the meander appears as an additional timing uncertainty (longitudinal geometric jitter $j_\mathrm{geom}$). The partial reflection from the termination to ground creates a local feature on the rising edge of the pulse (dashed circle), which acts as a partial compensation for the $j_\mathrm{geom}$. (b) The SNSPD is configured for an impedance-matched single-ended readout. The impedance matching taper interfaces to the $50\,\mathrm{\Omega}$ electronics, minimizes the reflections, and boosts the slew rate. The geometric effect of the transmission line is present, but the ground reflection still provides the local feature where $j_\mathrm{geom}$ is partially compensated. (c) The SNSPD is configured for an impedance-matched differential readout. Both ends of the nanowire are impedance matched and the complementary pulses (positive $V_\mathrm{pos}$ and negative $V_\mathrm{neg}$) experience no reflections. This setup allows the collection of the complementary time tags which can be post-processed to extract $t_\mathrm{\Sigma}$. The dashed box indicates the content shown in sub-figure (d). (d) A partial cancellation of the geometric contribution can be achieved by processing the difference of the complementary pulses, $V_\mathrm{diff}$. By optimizing the trigger level, $t_\Sigma$ can be directly tagged and the geometric contribution is minimized without the need for post-processing of the complementary time tags. In this case, differential readout enables partial compensation of the geometric jitter.}
	\label{fig:simulation}
\end{figure*}

To illustrate the distributed effects in SNSPDs, we consider a simple model resembling the architecture presented later on in this paper. A 100-\si{nm}-wide nanowire is arranged as a $20\,\si{\upmu m} \times 25\,\si{\upmu m}$ meander with $20\,\%$ fill-factor. The nanowire is assumed to have a sheet kinetic inductance of 80 \si{pH} per square (typical of $\approx 5\,\si{nm}$ thick NbN) and it is embedded in a cavity with a topside dielectric stack and a backside metallic gold mirror. See Appendix\,\ref{fab} for more details on the cavity material and design. In this environment, the nanowire behaves like a stripline with a characteristic-impedance $Z_0=3.26\,\si{k\Omega}$ and an effective microwave index $n_0=73.6$. The phase velocity is $v_\mathrm{ph}=4.1\,\si{\upmu m / ps}$ and a pulse takes $\approx 250\,\si{ps}$ to travel between the two ends of the detector. See Appendix \ref{app_microwave} for details on the microwave properties of nanowire stripline architectures.

In conventional readout designs, this SNSPD would be configured for single-ended readout with a $50\,\mathrm{\Omega}$ RF low-noise amplifier on one side and termination to ground on the other side. Figure\,\ref{fig:simulation}(a) shows the simulation of detection pulses out of the SNSPD, in this condition. The simulation includes both the electrothermal evolution of the hotspot and the microwave dynamics of the nanowire \cite{berggren2018superconducting}. See Appendix \ref{app_spice} for additional details on the simulations. The voltage pulses are characterized by several reflections and distortions caused by the impedance mismatch on both sides of the nanowire, which leads to a reduced slew rate and to a higher impact of the readout electrical noise on the timing jitter, referred to as amplifier jitter $j_\mathrm{amp}$ \cite{korzh2020demonstration,santavicca2019jitter}. The relative variance in the propagation delays generated by detection events in different areas of the meander (different curves in Fig.\,\ref{fig:simulation}(a)) appear as an additional uncertainty on the pulse time tag, referred to as longitudinal geometric jitter $j_\mathrm{geom}$\,\cite{calandri2016superconducting}.

For detectors large enough to couple to a single-mode fiber\,\cite{miller2011compact}, the geometric jitter can contribute on the order of tens of picoseconds to the total system jitter. Therefore, a trade-off between system detection efficiency and jitter must be typically made. In practice, as shown in the simulation of Fig.\,\ref{fig:simulation}(a), the partial reflection from the termination to ground creates a local feature on the rising edge of the pulse (dashed circle), which acts as a partial compensation for the $j_\mathrm{geom}$. The compensation feature on the rising edge was previously observed in Ref. \,\cite{marvinney2021waveform}, and described as a pulse-echoing effect due to impedance mismatch. Here, we recognize that this effect might explain the compensation of the geometric jitter in traditional detectors. Triggering at this optimal level produces a time tag with the geometric contribution partially compensated for, meaning the timing resolution can be significantly better than the worst-case scenario of the full propagation delay (250\,\si{ps} in this example). Nevertheless, this compensation feature is strongly dependent on other elements of the SNSPD design (e.g., pad layout, printed circuit board (PCB), ground termination) and does not always guarantee optimal timing resolution. Moreover, this treatment only applies for detectors in a distributed regime. In single-pixel designs with fully lumped behavior (delay-line effect can be neglected) \,\cite{esmaeilzadeh2020efficient,korzh2020demonstration}, the jitter is mainly limited by intrinsic contributions and electrical noise. 

The impact of the electrical noise $j_\mathrm{amp}$ can be mitigated by designing the SNSPD to preserve the integrity of the output pulse. To avoid the reflections caused by the impedance mismatch at the readout port, an impedance-matching taper interfacing the $\approx \mathrm{k\si{\Omega}}$ impedance nanowire to the $50\,\si{\Omega}$ readout can be integrated. Thanks to the impedance transformation, the output pulse has a higher amplitude and a faster slew rate, allowing a reduction of the electrical noise contribution to the timing jitter through a higher signal-to-noise ratio (SNR) \,\cite{zhu2019superconducting}. The simulation result in Fig. \ref{fig:simulation}(b) shows that the integration of a $200\,\si{MHz}$ cutoff Klopfenstein taper increases the output amplitude by a factor of 3.3, compared to the unmatched version in Fig.\,\ref{fig:simulation}(a). This intrinsic amplification is in agreement with previous experimental demonstration\,\cite{zhu2019superconducting}. The pulse rising edge is affected by just one reflection due to the ground termination. The geometric effect of the transmission line is present, but the ground reflection still provides the local feature where $j_\mathrm{geom}$ is partially compensated.

The geometric jitter can be compensated in an active fashion using a differential readout configuration\,\cite{calandri2016superconducting}. If the pulses coming from the two ends can be collected (positive pulse $V_\mathrm{pos}$ and negative pulse $V_\mathrm{neg}$), straightforward signal processing allows one to remove any dependence on the location of the hotspot along the nanowire, effectively reducing the timing jitter through the compensation of the geometric contribution. In fact, assuming the transmission line has a constant velocity $v_\mathrm{ph}$, the time tags from the two ends are $t_\mathrm{pos}=t_\mathrm{p}+\frac{x_\mathrm{p}}{v_\mathrm{ph}}$ and $t_\mathrm{neg}=t_\mathrm{p}+\frac{L-x_\mathrm{p}}{v_\mathrm{ph}}$, respectively, where $t_\mathrm{p}$ is the detection time and $x_\mathrm{p}$ is the hotspot location. The normalized sum of the time tags $t_\mathrm{\Sigma}=\frac{t_{\mathrm{pos}}+t_{\mathrm{neg}}}{2}=t_\mathrm{p}+\frac{L}{2v_\mathrm{ph}}$ is independent of the detection location and its jitter $j_\mathrm{\Sigma}$, will be unaffected by the geometric contribution.

In order to achieve the optimum geometric jitter cancellation and signal integrity, in this work, we use both a differential readout and impedance matching tapers on both ends of the detector. This design was inspired by the superconducting nanowire single-photon imager\,\cite{zhao2017single} as well as by early work in jitter reduction\,\cite{korzh2020demonstration,zhu2019superconducting} and differential readout\,\cite{calandri2016superconducting}, and the elements were specifically optimized for single-pixel SNSPDs. 

Figure\,\ref{fig:simulation}(c) shows the simulation results for a differential detector with matched readout. The complementary output pulses, $V_\mathrm{pos}$ and $V_\mathrm{neg}$, show no reflections and both have a superior slew rate compared to the unmatched pulse. To cancel the geometric jitter, one could post-process the time tags of the complementary pulses, which also makes it possible to gain information regarding the photon absorption location, something that will be demonstrated in the following sections. Equivalently, as shown in Fig.\,\ref{fig:simulation}(d), partial cancellation of the geometric contribution can be achieved by processing the difference of the complementary pulses $V_\mathrm{diff}$, if the photon-absorption location is not required. 

\subsection{Readout design: differential-to-single ended}\label{sec:readout}
\begin{figure}[!ht]
	\centering
	\includegraphics{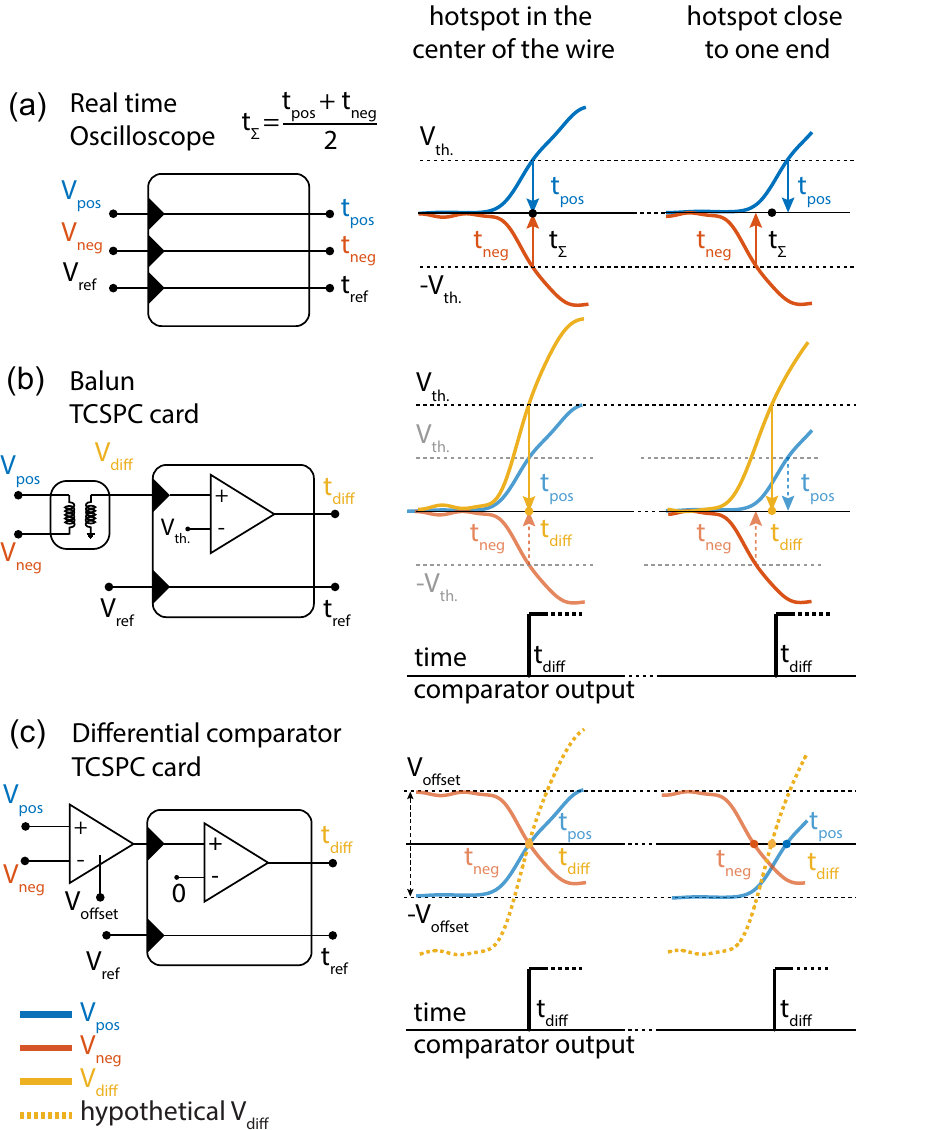}
	\caption{Sketch of various differential single-pixel readout setups and pulse processing. The pulses are shown for two arbitrary hotspot locations along the meander.  (a) A high-resolution real-time oscilloscope is used to collect the complementary pulses. The time tags $t_\mathrm{pos}$ and $t_\mathrm{neg}$, corresponding to the set threshold voltage $V_{\mathrm{th}}$, can be used to calculate $t_\Sigma$ (b) The complementary pulses are fed, after amplification, to the differential input of a balun that performs an analog difference of the pulses. The output is sent to the TCSPC module. If an appropriate threshold voltage is set, the TCSPC module will output a time tag $t_\mathrm{diff} \approx t_\mathrm{\Sigma}$. The jitter extracted with this method, $j_\mathrm{diff}$ represents the system jitter as it includes contribution from the readout system. (c) The complementary pulses are fed, after amplification, to the input of a differential comparator. To achieve optimum cancellation, a positive/negative offset is provided to the negative/positive pulse at the input of the differential comparator, respectively. When the difference between the two inputs becomes positive, the comparator generates a digital signal time tagged by the TCSPC module. If the appropriate offset is applied, the time tag $t_\mathrm{diff} \approx t_\Sigma$. To assist in understanding, the working principle of the comparator is illustrated with a dotted line, showing the $V_\mathrm{diff}$ between the two inputs. The jitter extracted with this method, $j_\mathrm{diff}$ represents the system jitter.}
	\label{fig:triggering}
\end{figure}

To achieve the best timing resolution we need a differential detector in combination with a high-resolution real-time oscilloscope, that served as a time tagger. The two ends of the SNSPD were directly fed to the input of the oscilloscope after amplification  (Fig. \ref{fig:triggering} (a)). The electrical noise jitter was minimized through the use of cryogenic RF amplifiers at each end. The trigger voltage $V_{\mathrm{th}}$ was set such as to minimize the noise contribution, by sampling the steepest point of the pulse\,\cite{korzh2020demonstration}. 

If the detection event happens at the exact center of the SNSPD (left side of the sub-figure), and the system is perfectly balanced, there will be no geometric contribution and $t_\mathrm{pos}=t_\mathrm{neg}=t_\Sigma$. When the detection event happens elsewhere (right side of the sub-figure), the pulses arrive with a relative delay induced by the transmission line effect. In this case, $t_\mathrm{\Sigma}$ can be processed and used to compensate for the geometric jitter contribution as illustrated in Fig. \ref{fig:triggering}(a). The jitter associated to $t_\mathrm{\Sigma}$, $j_\mathrm{\Sigma}$, represents the jitter of the detecting element alone, e.g. the detector jitter.

For practical single-photon counting applications, an instantaneous measurement is required, and the use of two low-jitter time taggers for each differential detector becomes impractical in many situations. To overcome this limitation,  we designed two readout schemes that make the differential detector compatible with a single ended TCSPC module while maintaining the advantages of differential compensation of the geometric jitter.  Figure\,\ref{fig:triggering} (b) and (c) shows a comparison of the two approaches. 

As previously shown in the simulation of Section\,\ref{sec_dev_sim}, compensation of the geometric jitter can be obtained by processing the difference of the complementary pulses $V_\mathrm{diff}$. In Fig.\,\ref{fig:triggering}\,(b) we illustrate how, instead of post-processing two individual time tags, $t_\mathrm{pos}$ and $t_\mathrm{neg}$, to obtain $t_\mathrm{\Sigma}$, a 2:1 balun transformer can be used to perform an equivalent analog difference of the complementary pulses. After amplification, the two sides of the detector are connected to the differential inputs of the balun, with the output being sent to a TCSPC module. The module will output a time tag $t_\mathrm{diff}$. The threshold voltage of the module can be set to minimize the spread of time tag distribution, corresponding to the condition $t_\mathrm{diff}  \approx t_\mathrm{\Sigma}$ (see Table\,\ref{tab:formulas} for an overview of the terminology).  The jitter extracted with this method, $j_\mathrm{diff}$, represents the timing resolution of the whole measurement system, the system jitter. Due to the insertion loss (6 \si{dB}) of the balun, we expect the system jitter to be slightly degraded compared to the detector jitter $j_\mathrm{\Sigma}$. 

Figure\,\ref{fig:triggering}\,(c) illustrates the use of a differential comparator, or equivalently, a differential-input TCSPC module. In this case, the differential comparator automatically cancels the geometric jitter, and no transformation of the pulses is required.  To achieve optimum cancellation, a positive/negative offset is provided to the negative/positive pulse at the input of the differential comparator. When the difference between the two inputs becomes positive, the comparator generates a digital signal, with a rising edge slope limited by the slew-rate of the comparator. This digital signal is time tagged by the TCSPC module producing the time tag $t_\mathrm{diff}$. The offset voltage can be set to minimize the spread of the time tag distribution, matching the condition $t_\mathrm{diff} \approx t_\Sigma$. This approach does not introduce insertion loss, thus it could be preferred over the 2:1 balun transformation, in certain situations.

\begin{table*}
	\centering
	\begin{tabular}{c|c|c|c|c|c|c}
		
		ID & type & wire width & active area/length & Cavity  $\lambda$ & NbN variant & $I_{sw}$  \\ \hline 
		A & meander & $100\,\nm$ & $25 \times 20\,\um^2$ &  $1550\,\nm$ & MIT & 14.6 \si{\mu A}\\
		B & meander & $100 \,\nm$ & $15 \times 10\,\um^2$  & $800\,\nm$  & JPL & 22 \si{\mu A}\\
		C & straight wire & $120\,\nm$ & $25 \,\um $ & $1550\,\nm$ & MIT& 22 \si{\mu A} \\ 
	\end{tabular}
	\caption{Characteristic of a subset of devices representative of the design space.}
	\label{tab:device}
\end{table*}
\section{Results} \label{sec_result}
In the following, we describe the experimental implementation of the approach proposed in Sec.\,\ref{sec_SNSPD_design}. In Sec.\,\ref{sec_result}(A), we describe the detector, analyze its output pulses, and discuss time tagging procedures. In Sec.\,\ref{sec_result}(B), we present the metrics achieved for the system detection efficiency and we correlate the values to the alignment of the fiber spot on the active area by exploiting the imaging capabilities of the detector, which we also introduce and treat theoretically.  In Sec.\,\ref{sec_result}(C), we first present and discuss the timing resolution metrics of the detector with a direct differential readout, using an oscilloscope. We then discuss the results achieved in combination with the differential-to-single-ended readout. In Sec.\,\ref{sec_result}(D), we present the photon-number resolution capabilities of our system. 

\begin{figure}[!ht]
	\centering
	\includegraphics{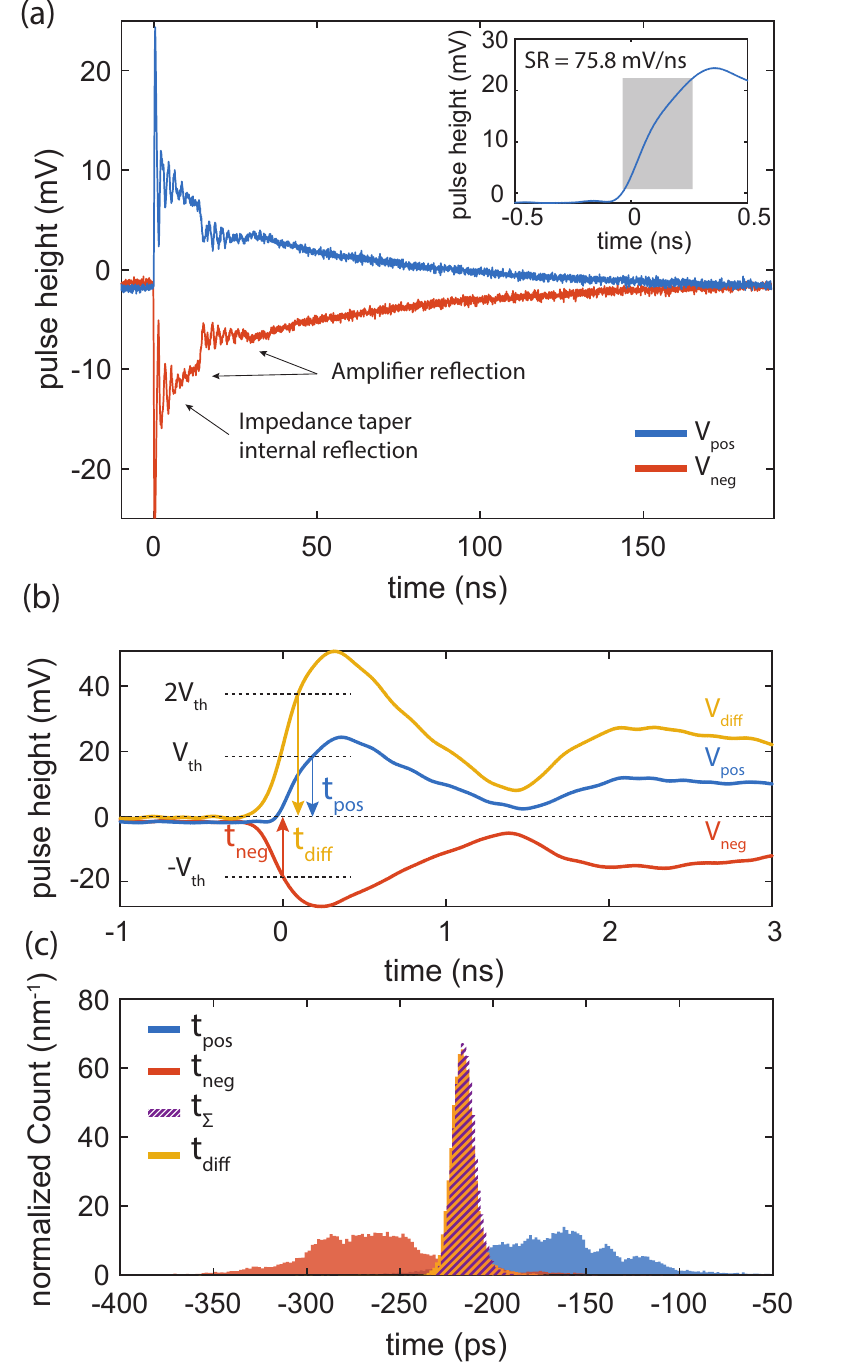}
	\caption{(a) Complementary pulses from device A upon illumination. The main signal discontinuities are attributed to reflections from the low-noise amplifier. The high frequency ripple is attributed to internal reflection in the impedance matching taper due to the limited design bandwidth. Inset: positive pulse slew rate. The impedance matching taper boosts the slew rate. (b) Time tagging procedure for $V_{\mathrm{pos}}$, $V_{\mathrm{neg}}$, and $V_{\mathrm{diff}}$. The complementary pulses generate two time tags $t_\mathrm{pos}$ and $t_\mathrm{neg}$, which are used to extract $t_\Sigma$ and estimate the detector jitter with the longitudinal geometric contribution compensated for. The difference of the pulses generates a time tag $t_\mathrm{diff}$, which is equivalent to $t_\Sigma$. (c) Time tag distributions for $123348$ detection events. $t_\mathrm{diff}$-distribution is substantially narrower than the individual distributions of $t_\mathrm{pos}$ and $t_\mathrm{neg}$ and is approximately equivalent to the $t_\mathrm{\Sigma}$-distribution.}
	\label{fig:impedance_matched_signal}
\end{figure}
\subsection{Impedance-matched differential detectors}

Our impedance-matched differential designs were realized with niobium nitride (NbN). Optical micrographs of the detectors are shown in Fig.\,\ref{fig:architecture}(a). Details of the fabrication process are reported in Appendix \ref{fab}. We characterized SNSPDs from several fabrication runs, each with different designs and specifications, as well as two variants of niobium nitride. The variants are identified with the institution where the deposition was performed, i.e. at the Massachusetts Institute of Technology (MIT) or at the Jet Propulsion Laboratory (JPL). In Table \ref{tab:device} we summarize the main characteristics of a selected subset of detectors, representative of the wide design-parameter space. 

All the reported detectors reached saturation of the internal detection efficiency and were operated on the system detection efficiency plateau. The detectors with MIT-grown NbN had slightly lower switching current ($I_\mathrm{sw}$) compared to JPL-grown NbN, for same nominal thickness. Overall, the detector switching currents ranged between  $14\,\mathrm{\mu A}$ and $25\,\mathrm{\mu A}$.

Figure\,\ref{fig:impedance_matched_signal}(a) shows the complementary pulses from Device A (see Table\,\ref{tab:device}), biased with 14\,\si{\mu A} (96\,\% of $I_\mathrm{sw}$), and illuminated with a laser having a central wavelength of 1550 \si{nm}, a pulse width of $1\,\si{ps}$, and a repetition rate of $10\,\si{MHz}$. The output pulses were amplified with two, high dynamic range, $2\,\mathrm{GHz}$ cryogenic low-noise amplifiers (see Appendix\,\ref{cryoamp}) and acquired with an oscilloscope with a sample rate of $80 \times 10^9 \,\mathrm{s^{-1}}$. The slew rate SR is 75.8 \si{mV/ns} (inset) and the reset time is $\approx 160\, \si{ns}$. The high-frequency ripple visible up to $\approx 50\,\si{ns}$ is due to the internal reflections in the matching tapers, while the reflections located between $\approx 14\,\si{ns}$ and $\approx 30\,\si{ns}$ are due to negative reflections from the low-noise amplifiers. 

Figure\,\ref{fig:impedance_matched_signal}(b) shows the time tagging procedure for the complementary pulses as well as the timing of their difference, $V_{\mathrm{diff}}$. Note that for $V_\mathrm{diff}$ we used a threshold voltage of twice the positive pulse threshold. The complementary pulses generate two time tags $t_\mathrm{pos}$ and $t_\mathrm{neg}$, which are post-processed to extract $t_\Sigma$  and estimate the detector jitter $j_\Sigma$ (see Table\,\ref{tab:formulas})  with the longitudinal geometric contribution compensated for. To demonstrate that the method described in Sec.\,\ref{sec:readout}, of using the difference of the pulses to generate a time tag $t_\mathrm{diff}$, is equivalent to the calculated $t_\Sigma$, we implement the operation in post-processing on the analog waveforms. As shown in Fig.\,\ref{fig:impedance_matched_signal}(c) the $t_\mathrm{diff}$-distribution calculated for more than one hundred thousands detection events is substantially narrower than the individual distributions of $t_\mathrm{pos}$ and $t_\mathrm{neg}$ and is approximately equivalent to the $t_\mathrm{\Sigma}$-distribution. This demonstrates that the operation can be implemented with analog components as proposed in Sec.\,\ref{sec:readout} and will be further discussed in Sec.\,\ref{sec:jitter}.

\subsection{System detection efficiency and delay line imaging capabilities}

\begin{figure}[!ht]
	\centering
	\includegraphics{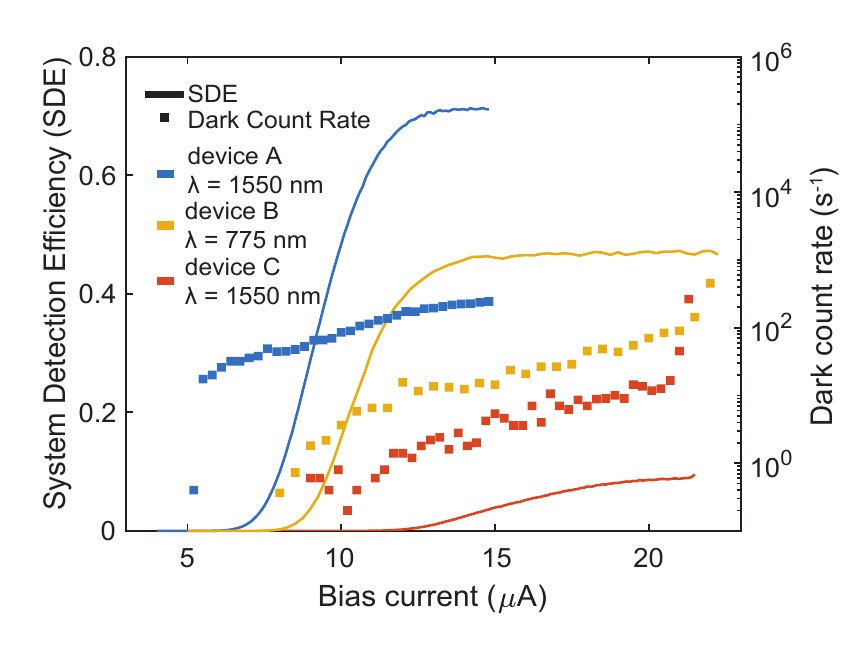}
	\caption{System detection efficiency curves for detectors A, B, and C. Solid lines indicate the system detection efficiency, the markers indicate the dark count rate.  All the detectors achieved saturation of the internal detection efficiency.}
	\label{fig:diff_jitter}
	
\end{figure}
Figure\,\ref{fig:diff_jitter} shows the system detection efficiency curves obtained for the device selected for this paper. Detector A achieved 71\% saturated system detection efficiency with $\approx 200\,\si{s^{-1}}$ dark count rate. The efficiency is limited by the detector fill-factor (20\%) and the absorption in the metallic mirror. Further investigations into the trade-off of the fill factor and microwave propagation in the active region of the differential SNSPD, as well as the use of a more reflective metal for the mirror/ground layer, may lead to an increase of the system detection efficiency in the future. Detector B achieved $47.6\%$ saturated detection efficiency at $775\,\si{nm}$ with a dark count rate of $\approx 100\,\si{s^{-1}}$. The single-wire geometry of detector C determines a $8.8\%$ saturated detection efficiency at 1550\,\si{nm} with a dark count rate of $\approx 10\,\si{s^{-1}}$.
The relative uncertainty on the all system detection efficiency values is approximately $5.2\%$. See Appendix \ref{app_uncertainty} for more details on this estimation. 
To explain the discrepancies between the values of detection efficiency between the detector A and B we investigated the quality of the fiber alignment by exploiting the delay-line imaging capabilities of our detectors. 
 
\begin{figure*}
	\centering
	\includegraphics{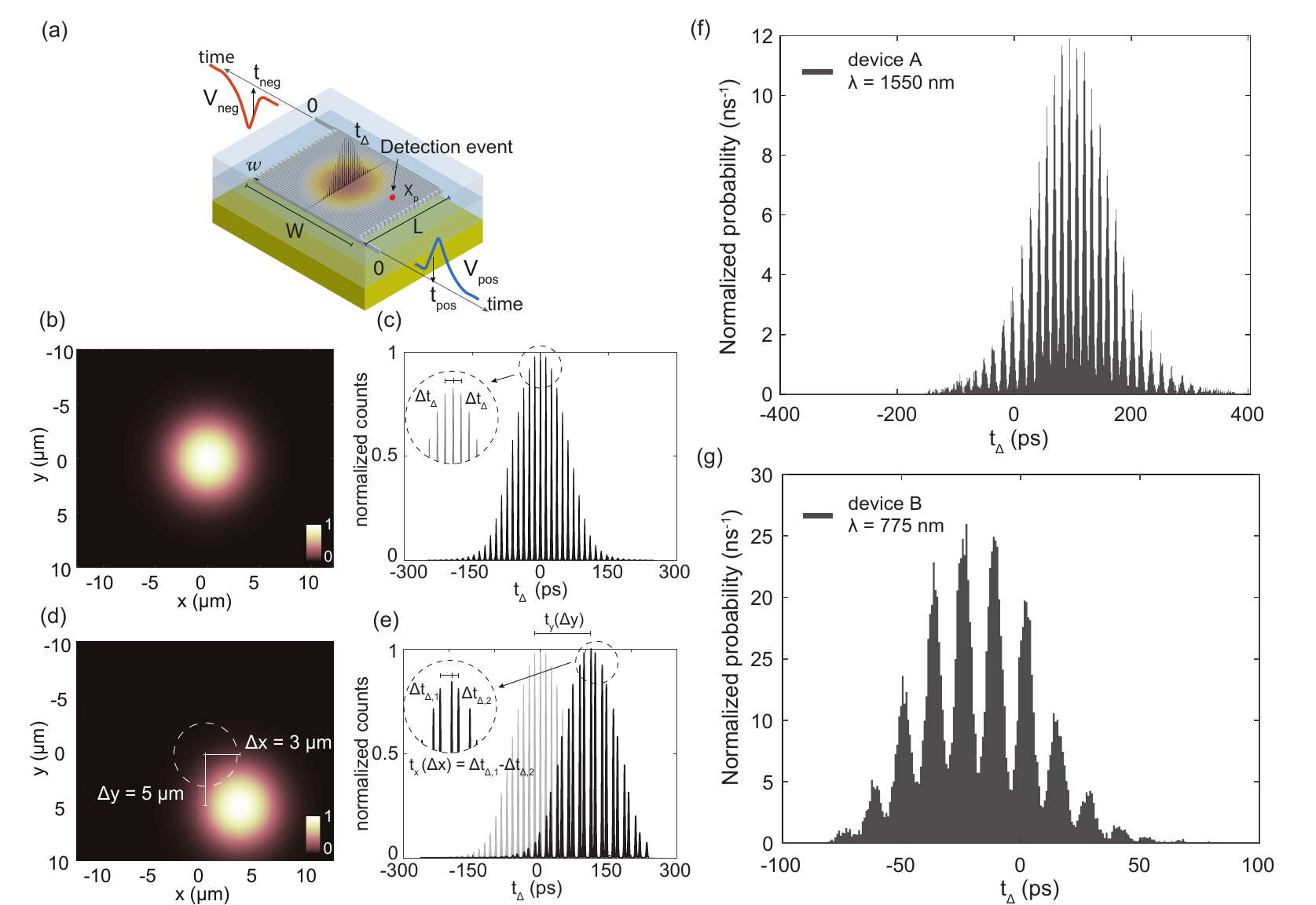}
	\caption{Single-mode differential delay-line imaging. When a differential readout is utilized and the speed of signal propagation on the order of few $\si{\upmu m /ps}$, differential detectors can be used for localizing a light beam. (a) A single LP01 mode illuminates the center of a $L \times W$-area meander with $w$-wide nanowire. The coordinate of the detection event $x_\mathrm{p}$ can be determined by collecting both of the time tags $t_\mathrm{pos}$ and $t_\mathrm{neg}$ from the differential readout setup. (b) LP01 mode is aligned with the detector center. (c) Simulation of the differential time $t_\mathrm{\Delta}$-histogram. Each sub-distribution corresponds to detection event in single wires of the meander.  (d) LP01 mode and the meander centers are misaligned by $\Delta x = 3\,\si{\upmu m}$ and $\Delta y = 5\,\si{\upmu m}$. (e) Simulation of the differential time $t_\mathrm{\Delta}$-histogram. When the mode is misaligned the histogram shows two peculiarities. The shift of the center $t_\mathrm{y}$ encodes the vertical misalignment $\Delta y$, while the difference between relative spacing of adjacent peaks $t_\mathrm{x}=\Delta t_\mathrm{\Delta,1}-\Delta t_\mathrm{\Delta,2}$ encodes the horizontal misalignment $\Delta x$. (f) Differential time distribution $t_\mathrm{\Delta}$ detector A. (g) Differential time distribution $t_\mathrm{\Delta}$ detector B. }
	\label{fig:single_mode_imaging}
\end{figure*}

With a differential readout \cite{calandri2016superconducting,zhao2017single} the relative delay between the time tags of the complementary pulses $t_\mathrm{\Delta} =t_{\mathrm{pos}}-t_{\mathrm{neg}}=\frac{L_\mathrm{m}-2x_{\mathrm{p}}}{v_{\mathrm{ph}}}$ encodes the spatial coordinate of the photon detection location $x_\mathrm{p}$ on the nanowire (Fig. \ref{fig:single_mode_imaging}(a)); $L_\mathrm{m}$ is the total length of the meander. In differential time tag multiplexed SNSPD arrays and single-photon imagers \cite{zhao2017single,zhu2018scalable} the pixels are separated by a fixed length of delay line or are part of a continuous delay line. The $t_\mathrm{\Delta}$-histogram is used to determine the spatial distribution of the photon counts and reconstruct the image.

Thanks to the stripline-like design of our differential detector, the signal propagation velocity is low enough to enable the use of $t_\mathrm{\Delta}$ to precisely determine the photon absorption locations along the nanowire meander. For the same detector model as in the previous section, the readout signal travels at a velocity of $4.1\,\si{\upmu m/ps}$ in the nanowire, meaning that it takes approximately $6.1\,\si{ps}$ for the signal to traverse a single meander. This delay can be resolved with high-resolution time taggers %and our single-pixel detector achieves imaging capability by time-multiplexing adjacent locations on the meander.
Moreover, the monolithic impedance-matching and the use of a cryogenic amplifier preserves the pulse leading edge, avoids distortions, and minimizes the added electrical noise, which preserves the original timing information.

To showcase this method, we first simulated the imaging capability of the detector for the lowest-order mode LP01 of a single-mode fiber at $1550\,\si{nm}$. The mode is assumed to land on the SNSPD without spreading or distortions, maintaining the same size as at the end of the fiber. Figure\, \ref{fig:single_mode_imaging}(c) shows the simulated distribution of the differential time $t_\mathrm{\Delta}$ when the center of the optical mode coincides with the center of the detector (Fig. \ref{fig:single_mode_imaging}(b)). Each sub-distribution of the histogram corresponds to detection events from consecutive wires in the meander. Because the mode is aligned with the center of the meander, the spacing between adjacent sub-distributions is constant, and the peak of the envelope of the overall $t_\Delta$-distribution ($t_\mathrm{y}$) is located at $0\,\mathrm{ps}$. 

When the mode and the detector centers are misaligned (e.g. $\Delta x=3\,\si{\upmu m}$ and $\Delta y=5\,\si{\upmu m}$ in Fig. \ref{fig:single_mode_imaging}(d)) the histogram shows two characteristic features (Fig. \ref{fig:single_mode_imaging}(e)). First, $t_\mathrm{y}$ does not coincide with $t_\mathrm{\Delta}=0$. In fact $t_\mathrm{y}$ encodes the vertical shift relative to the detector center according to $\Delta y = t_\mathrm{y}\frac{v_\mathrm{ph}}{2} \frac{w}{W\,\mathrm{FF}}$. Here, $w$ is the width of the nanowire width, $W$ the width of the meander active area, $L$ is the length of the meander active area, and $\mathrm{FF}$ is fill-factor of the meander. For clarity, the variables are also defined pictorially in Fig. \ref{fig:single_mode_imaging}(a). Second, the relative spacings between adjacent peaks, $\Delta t_\mathrm{\Delta,1}$ and $\Delta t_\mathrm{\Delta,2}$, are not identical.  In fact, their difference $t_\mathrm{x}(\Delta x)=\Delta t_\mathrm{\Delta,1}-\Delta t_\mathrm{\Delta,2}$ encodes the horizontal offset of the events from the center of the meander according to $\Delta x=\frac{v_\mathrm{ph}t_\mathrm{x}}{8}$. See Appendix \ref{form_dev} for the derivation of these first order formulas. In this specific case, $t_\mathrm{y}(5\,\si{\upmu m})=118.3\,\si{ps}$ and $t_\mathrm{x}(3\,\si{\upmu m})=5.8\,\si{ps}$.

When the mode is unknown, the difference time histogram can be used to reconstruct the spatial distribution of the light and identify the nature of the source and illumination conditions. 
When the mode is known, analysis of the $t_\mathrm{\Delta}$-histogram is specifically useful to debug fiber coupling alignment, assess fabrication yield, estimate upper bounds for detection efficiency, and determine the effective phase velocity of the detector. Figure\,\ref{fig:single_mode_imaging}(f) and (e) show direct exploitation of this method to check for mode alignment with the detector active area. 

Figure\,\ref{fig:single_mode_imaging}(f) shows the $t_\mathrm{\Delta}$-distribution for Detector A illuminated with the $1550\,\si{nm}$ pulsed laser fiber coupled to the detector through a single mode fiber. We fitted the distribution with a gaussian model to estimate the mode misalignment. For detector A, the misalignment is estimated to be $\Delta x = -0.01\,\si{\mu m}$ and $\Delta y = 3.74\,\si{\mu m}$. The fraction of the mode collected by the active area was $99.7\,\%$. This analysis confirms that the SDE is limited to $\approx 70\%$ at $1550\,\mathrm{nm}$ by the meander fill-factor and cavity design.

Figure\,\ref{fig:single_mode_imaging}(g) shows the the $t_\mathrm{\Delta}$-distribution for Detector B illuminated with the $775\,\si{nm}$ pulsed laser fiber coupled to the detector through a single mode fiber. In this case, inspection of the histogram reveals that the mode was strongly misaligned toward the lower corner of the meander. Combined with the smaller active area, the detector can only reach $\approx 47\%$ SDE. When aligned, the system detection efficiency should exceed $70\,\%$, based on Detector A characterization.

\subsection{Detector jitter, system jitter and FW1/100M}\label{sec:jitter}

\begin{figure*}
	\centering
	\includegraphics{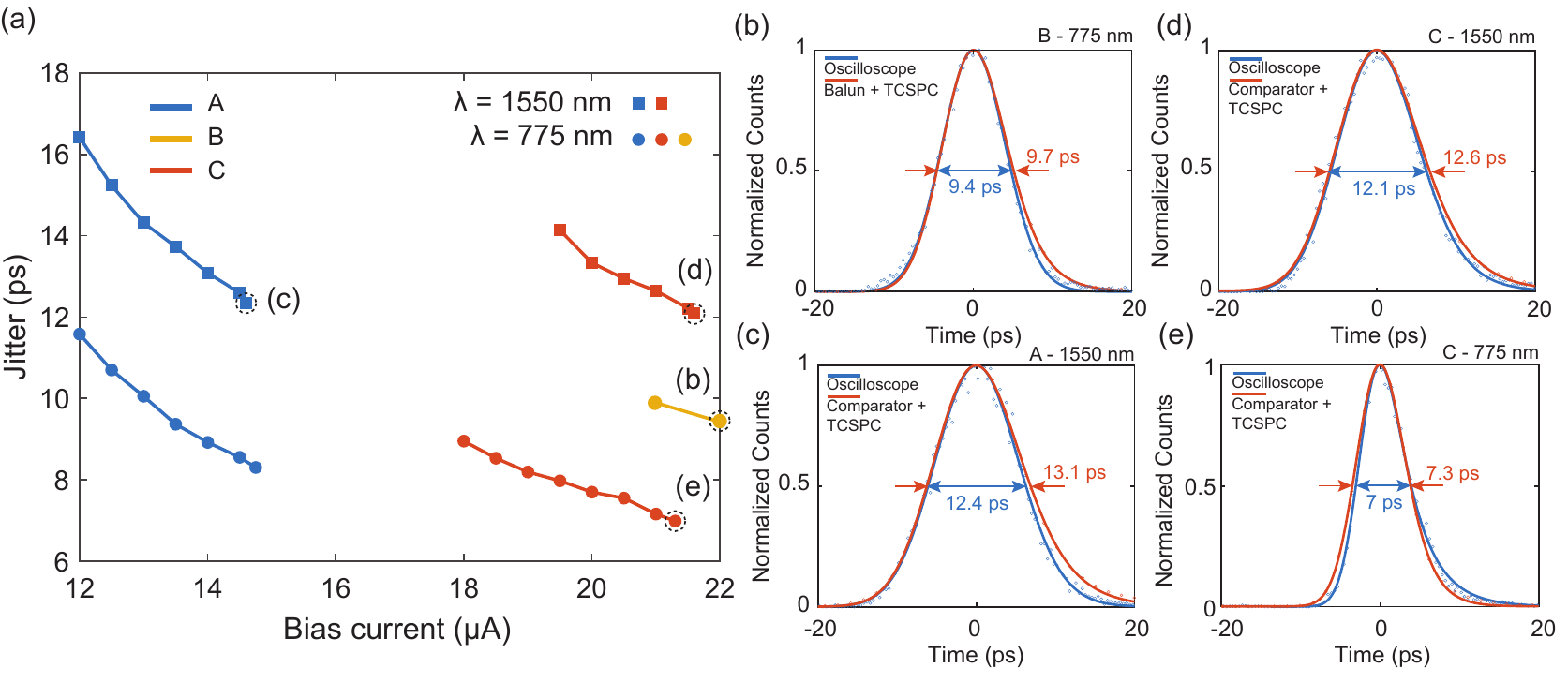}
	\caption{Jitter. (a) Dependence of the detector jitter, $j_{\mathrm{\Sigma}}$, across several detectors designs, wavelengths, and bias current. Both trends indicate that the dominant contribution is intrinsic. (b-e) Histograms corresponding to the points indicated in sub-figure (a). Comparison of detector jitter and system jitter obtained with the methods described in the paper. The system jitter obtained with the differential-to-single-ended readout is at most 6\% higher compared to the detector jitter, showing that the additional readout electronics has a minimal impact on the overall timing resolution. }
	\label{fig:comparison_jitter}
\end{figure*}
\begin{figure}[!ht]
	\centering
	\includegraphics{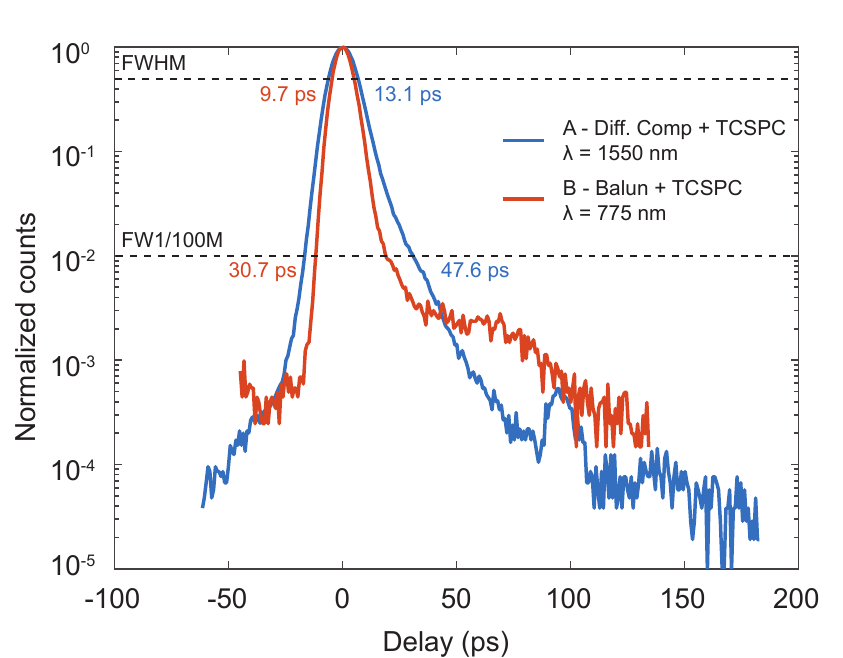}
	\caption{Timing response for detector A at $\lambda=1550\,\si{nm}$, in combination with the differential comparator and TCSPC module, and for detector B at $\lambda=775\,\si{nm}$, in combination with the balun  and TCSPC module.}
	\label{fig:fw1/100M}
\end{figure}
Figure\,\ref{fig:comparison_jitter}(a) shows the detector jitter $j_\mathrm{\Sigma}$ and its dependence on bias current and wavelength for the three detectors listed in Table\,\ref{tab:device}. The jitter was obtained as the full-width at half maximum (FWHM) of the exponentially-modified Gaussian function used to fit the $t_\mathrm{\Sigma}$-distribution. The uncertainties are calculated as the $95\,\%$ confidence bounds on the data. $t_\mathrm{\Sigma}$ was calculated with the time tags of the complementary pulses acquired with an oscilloscope with a sample rate of $80 \times 10^9 \,\mathrm{s^{-1}}$. The jitter decreases for the shorter wavelength and for bias currents closer to the switching current, which indicates that the dominant contribution is the intrinsic jitter of the detector\,\cite{korzh2020demonstration, allmaras2019intrinsic}, demonstrating that the detector design succeeds in overcoming the effects of the geometric and amplifier jitter to enable scalability to practical active areas. Detector A achieves a detector jitter of $12.4 \pm 0.6 \,\si{ps}$ at the target wavelength of $1550\,\mathrm{nm}$ and a detector jitter of $8.3\pm 0.3 \,\si{ps}$ at 775\,\si{nm}. Detector B achieves a detector jitter of $9.4 \pm 0.8 \,\si{ps} $ at its target wavelength of $775\,\mathrm{nm}$. Detector C achieves a detector jitter of $12.1 \pm 0.5 \,\si{ps} $ at $1550\,\mathrm{nm}$. For $775\,\mathrm{nm}$ photons, it achieves the lowest detector jitter of $7.0 \pm 0.3 \,\si{ps}$, thanks to its straight wire geometry, with minimal geometric contribution by design and lowest probability of film defects, enabling operation at a higher fraction of the depairing current\,\cite{frasca2019determining}. We expect that the intrinsic jitter of the straight wire geometry can be improved even further through the use of narrower wires (here $\approx 120\,\mathrm{nm}$)\,\cite{korzh2020demonstration}.

Figure\,\ref{fig:comparison_jitter}(b) compares the distribution of $t_\mathrm{\Sigma}$ with $t_\mathrm{diff}$ obtained with the methods described in Sec \ref{sec:readout} using a TCSPC module (Becker \& Hickl SPC-150NXX \footnote{Certain commercial equipment, instruments, or materials are identified in this paper to facilitate understanding. Such identification does not imply recommendation or endorsement by NIST, nor does it imply that the materials or equipment that are identified are necessarily the best available for the purpose.}) in combination with a balun or a cryogenic differential comparator (see Appendix\,\ref{cryoamp}). The proposed acquisition schemes achieve a $j_\mathrm{diff}$, of only $3\,\%$ to $6\,\%$ higher than $j_\mathrm{\Sigma}$, demonstrating that both methods achieve effective cancellation of the geometric jitter equivalent to the oscilloscope-based acquisition, while minimally affecting the overall timing resolution. The uncertainty on these values are mainly due to the resolution of the TCSPC module and estimated as two time-bins $\sigma_\mathrm{TCSPC}=0.4\,\mathrm{ps}$.  This opens up the possibility of using the detection system for photon-counting applications with high detection efficiency and sub-10\,ps system jitter, while operating at count rates in the MHz range, something that is not possible with oscilloscope-based data acquisition and that has not been achieved previously for the wavelengths in question.

In applications such as quantum key distribution or pulse-position modulated optical links, in order to achieve a low error rate and a high clock rate\,\cite{boaron2018secure,grunenfelder2020performance, amri2016temporal}, an instrument response function with a few-ps spread over several order of magnitudes (high dynamic range) is required. In addition to a low timing jitter (FWHM of the timing response), this characteristic is quantified by the full-width at one hundredth-of-maximum (FW1/100M) of the instrument response function. In fluorescent lifetime imaging, a large FW1/100M can limit the contrast, while in time-resolved spectroscopy, the dynamic range is affected\,\cite{hsu2009cmos,sanzaro2017single}.   
Figure\,\ref{fig:fw1/100M} shows that our differential Detector A, in combination with the differential comparator and the TCSPC module, achieves $47.6\,\si{ps}\pm 0.4\,\si{ps}$ FW1/100M at 1550\,\si{nm}, which is a factor of four lower than what has been achieved with free-running InGaAs/InP single photon avalanche diodes (SPAD), operating at the same wavelength\,\cite{amri2016temporal}. At 775\,\si{nm}, Detector B combined with the balun and TCSPC module, achieves $30.7\,\si{ps}\pm 0.4\,\si{ps}$ FW1/100M, which is a factor of seven lower than the best demonstration with red-enhanced silicon SPADs\,\cite{sanzaro2017single}. These metrics position our differential detector for application in biomedical imaging\,\cite{bruschini2019single, sutin2016time}, quantum communication\,\cite{hadfield2009single} and laser ranging\,\cite{mccarthy2013kilometre}, where the most stringent timing performance is required over a large dynamic range.

\subsection{Photon number resolution}

Recently, it was demonstrated that the output pulse amplitude of impedance-matched tapered SNSPDs can directly encode the number of photons detected simultaneously \cite{zhu2020resolving}. In the original STaND (Superconducting Tapered Nanowire Detector), the quasi-lumped nature of the photon sensitive area, and the single-ended readout, make the pulse amplitude scale sublinearly with the photon-number-dependent hotspot resistance $R_{\mathrm{HS}}(n)$. Here, the same simple picture can be applied, but the differential character of the detector encodes the photon-number information in the difference of the pulses from the two ends, $V_\mathrm{diff}$.

\begin{figure}[!ht]
	\centering
	\includegraphics{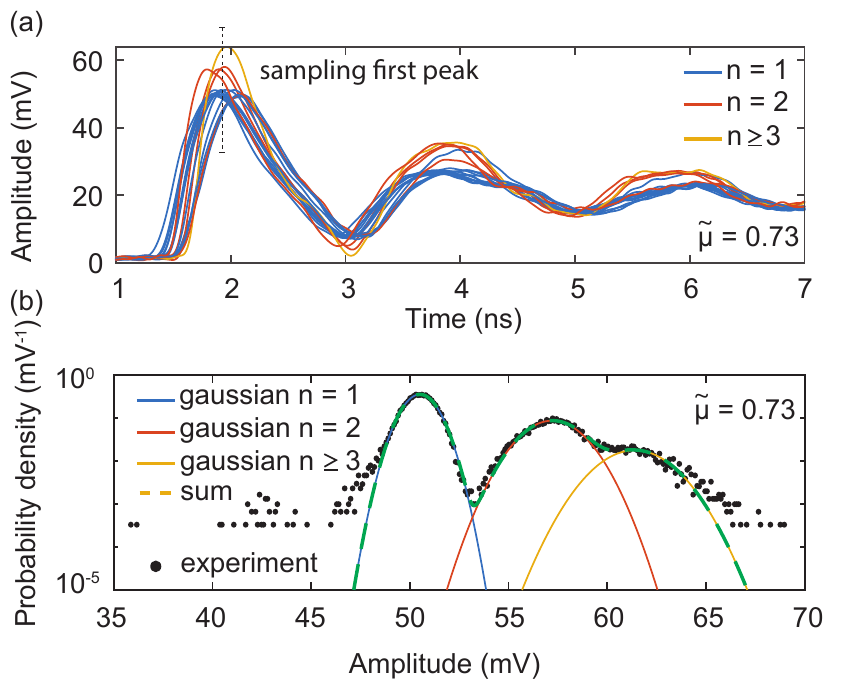}
	\caption{Detector output and counting statistics under coherent state illumination. (a) Sample of detector output $V_\mathrm{diff}$ for a pulsed laser with $\tilde{\mu} \approx 0.73$. The pulses have been grouped and colored by pulse height. (b) Gaussian fitting of the pulse amplitude histograms for $\tilde{\mu} \approx 0.73$, sampling the first peak (dashed line in (a)).}
    \label{fig:PNR}
\end{figure}

We characterized the photon-number discrimination capability using an attenuated $1550\,\mathrm{nm}$ pulsed laser with a repetition rate of $1\,\mathrm{MHz}$ (see Appendix \ref{app_setup} information for details of measurement setup). Figure\,\ref{fig:PNR} (a) shows a sample of representative traces of the difference of the output pulses for an effective mean photon number $\tilde{\mu}=0.73$. $\tilde{\mu}$ was estimated from the photon rate at the cryostat input port, scaled by the system efficiency (see Appendix \ref{app_pnr}). The pulse amplitude distribution sampled on the the first peak (dashed line in sub-figure (a)) is shown in Fig.\,\ref{fig:PNR}(b). Our detector can distinguish up to $n=3$ photons. We fitted the distribution using three Gaussian functions representing the number of photons. The separation between the one- and two-photon distributions is more than 9 standard deviations of the one-photon distribution width $\sigma_{n=1}$ ($9\sigma_{n=1}$), making this detector suitable for application in quantum optics experiments. Note that transmission line effects cause the amplitude and the shape of the differential output pulses to depend on the photon landing location. The effect tends to increase the width of the amplitude distribution, with a larger impact for multi-photon events, where the relative distance between landing locations plays an additional role (see Appendix \ref{app_pnr}). The additional features in the pulse (second and third peaks) might provide additional information and allow the discrimination of higher photon numbers through signal processing.

\section{Discussion}
\label{sec_discussion}
In this paper, we demonstrated that by redesigning the architecture of the SNSPD and with an appropriate readout scheme, the trade-off between high detection-efficiency and low system-jitter can be overcome. The impedance-matched design enables photon-number resolution capabilities and the differential design enables intrinsic device-level imaging capabilities.

A current limitation in the performance of the current design is the maximum count rate. Although we did not perform specific measurements to characterize this metric, we expect the count rate to be ultimately limited by the reset time of our detectors, approximately $160\,\mathrm{ns}$. To address this issue, an active quenching circuit could be coupled to the device\,\cite{ravindran2020active} and integrated on chip in future iterations. 

The design approach we provided in this paper traces a path to the realization of detectors with high system detection efficiency and high timing resolution. Nevertheless, in the current demonstration the performances are still below what could be achieved with designs focus on a single metric. For the system detection efficiency, assuming an optimal fiber alignment, the current limitations are attributed to the detector fill-factor and the cavity design. Both design elements were solely selected to facilitate fabrication and design and could be improved in future iterations.

The detector jitter we measured for our devices approached the few-ps domain. Nevertheless, they are still two to three times higher than the values obtained with specialized low-jitter devices. As discussed above, in our SNSPDs the detector jitter trends with respect to bias current and wavelength suggest that the dominant jitter contribution has intrinsic nature. Compared to the current record detector-jitter device \cite{korzh2020demonstration} ($4.3\,\mathrm{ps}$ at $1550\,\mathrm{nm}$), device A and B feature a larger areas. While the differential design effectively cancels the geometric contribution, large active areas devices can be affected with a higher probability by defects induced by nanofabrication (e.g. line edge roughness) or intrinsic to the film (e.g. natural constrictions and grain boundaries). In fact, the record jitter device was operating at a 0.8 fraction of the critical depairing current ($I_\mathrm{dep}$) while, based on previous measurements \cite{frasca2019determining}, we expect our device to operate between $0.5 I_\mathrm{dep}$ to $0.7I_\mathrm{dep}$. This explains the overall higher jitter values obtained in our meandered SNSPDs. Device C, although in a straight wire configuration, has a switching current considerably suppressed compared to the reference record device. This could explain the higher jitter. In this demonstration, we did not explicitly focused on optimizing the fabrication and film quality. We expect that by improving these technical aspects in the future we will reach timing resolutions closer to the current record values.  

The differential-to-single-ended setups rely on using external components (balun or differential comparator), either at room or cryogenic temperature. Although our experiments shows that the system performances are minimally degraded compared to the detector performances, external components make the detector prone to added electrical noise, which can ultimately degrade the system jitter. To make the system more compact and improve the performance, similar electrical circuits could be custom designed, integrated and co-located on-chip with the differential impedance-matched detector. Moreover, niobium nitride nanowires are also used in cryoelectronics applications (e.g. x-Tron family \cite{mccaughan2014superconducting,mccaughan2016using}) and were recently exploited to realize compact microwave devices\,\cite{colangelo2020compact,wagner2019demonstration}. Design and monolithic integration of nanowire-based electronics for on-chip signal conditioning could significantly improve the system performances and avoid additional post-processing (e.g. photon number discrimination). 

\section{Summary and conclusion} \label{sec_outlook}
Our differential impedance-matched detector, coupled with high-speed differential-to-single-ended readout, achieves high system detection efficiency and low system jitter while maintaining the convenience of single-ended readout. At $775\,\mathrm{nm}$ we achieved $9.7\,\si{ps}\pm 0.4\,\si{ps}$ FWHM and $30.7\,\si{ps}\pm 0.4\,\si{ps}$ FW1/100M system jitter with $ 47.3 \,\% \pm 2.4\,\%$ system detection, limited by fiber alignment. With a optimal alignment, this detector can achieve an SDE $> 70\,\%$. At 1550\,$\mathrm{nm}$ we achieved $13.1\,\si{ps}\pm 0.4\,\si{ps}$ FWHM and $47.6\,\si{ps}\pm 0.4\,\si{ps}$ FW1/100M system jitter with $71.1\,\% \pm 3.7\,\%$ system detection efficiency. 

These performance may enable quantum communication at clock-rates $>$20\,GHz, high-resolution single-photon laser ranging, faint optical-waveform reconstruction and previously unachievable capabilities in biomedical imaging applications.
The microwave design of our nanowire detectors combined with the differential readout architecture enables precise discrimination of the photon absorption locations along the nanowire meander, unlocking delay-line imaging capabilities.  We exploited this property to image the fiber mode projected on the detector and verify the fiber alignment.
Our detectors also achieve photon number resolution up to three photons. The possibility of discriminating the number of photons from optical radiation with high efficiency and timing resolution may enable the use of our detectors in applications such as non-classical state generation, quantum information processing, and linear optical quantum computing. 
The combined properties and achieved performance make our detector a versatile photon-counting solution for several applications.

\begin{acknowledgments}
Research was sponsored in part by the Army Research Office (ARO) and was accomplished under Cooperative Agreement Number W911NF-16-2-0192 and W911NF-21-2-0041. The views and conclusions contained in this document are those of the authors and should not be interpreted as representing the official policies, either expressed or implied, of the Army Research Office or the U.S. Government. The U.S. Government is authorized to reproduce and distribute reprints for Government purposes notwithstanding any copyright notation herein. Support for this work was provided in part by the Defense Advanced Research Projects Agency (DARPA) Defense Sciences Office (DSO) DETECT and Invisible  Headlights programs, the NASA Spacecraft Communication and Navigation (SCaN) technology development program, the Alliance for Quantum Technologies’ (AQT) Intelligent Quantum Networks and Technologies (INQNET) research program, and the National Science Foundation (NSF) grant under contract No. ECCS102000743. D.Z. acknowledges support by the National Science Scholarship from A*STAR, Singapore, and Harvard Quantum Initiative Postdoctoral Fellowship. J.P.A. acknowledges support from a NASA Space Technology Research (NSTRF) fellowship. A.B.W. acknowledges support from the NASA Postdoctoral Program at the Jet Propulsion Laboratory. Part of  this research  was performed  at the Jet  Propulsion Laboratory, California Institute of Technology, under  contract  with  NASA. The authors thank S. Weinreb and J. Bardin for useful technical discussions. The authors thank J. Daley and M. K. Mondol of the MIT Nanostructures Laboratory Facility for technical support. The authors thank P. Keathley, M. Bionta, and A. Bechhofer for assistance in editing the final manuscript.

\end{acknowledgments}

\section*{Authors Contribution}

M.C. and B.K. conceived, designed, and performed the experiments. M.C., A.D.B., B.B. and R.M.B. fabricated the devices. M.C. performed the simulations. M.C., B.K., and J.P.A. analyzed the data. M.C., A.D.B., B.K., J.P.A., A.S.M., R.M.B, B.B., M.R., M.J.S., A.N.M., D.Z., S.S., W.B., L.N., J.C.B., S.F., A.E.V., C.P., E.E.R., A.B.W., E.S., E.E.W., M.S., R.M., S.W.N. contributed materials/analysis tools. K.K.B., M.D.S. supervised the project. M. C. and B. K. wrote the paper with inputs from all authors.

\appendix
\section{Microwave Design}
\label{app_microwave}
\begin{figure}[!ht]
    \centering
    \includegraphics[width=\linewidth]{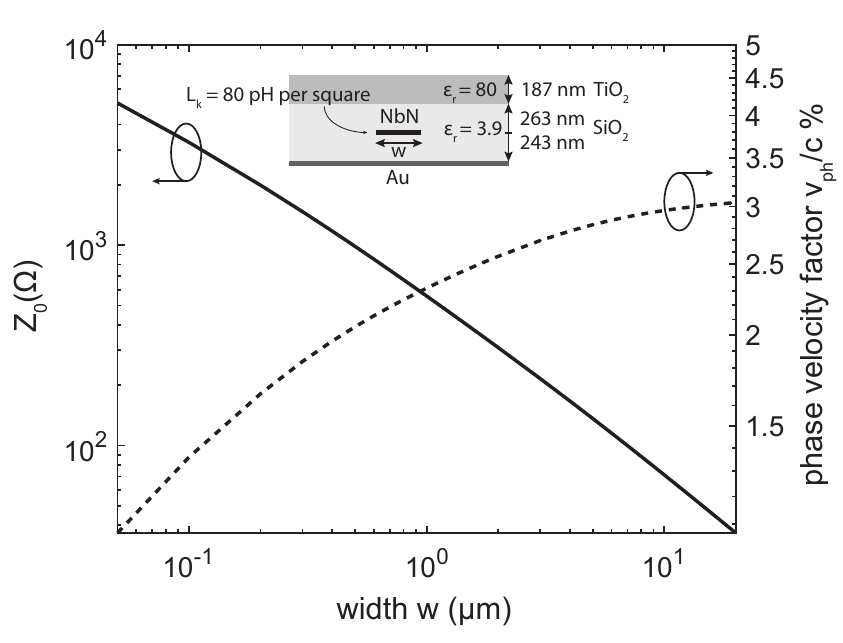}
    \caption{Simulation of the nanowire stripline parameters:   characteristic impedance and phase velocity factor versus conductor width for a sheet kinetic inductance $L_\mathrm{k}=80\,\mathrm{pH}$ per square.}
    \label{supfig:microwave}
\end{figure}

In this section we discuss the microwave design for our detector optimized for $1550\,\mathrm{nm}$ radiation. The superconducting nanowire is embedded in an $\mathrm{SiO_2}/\mathrm{TiO_2}$ optical cavity with a bottom metallic reflector. The silicon oxide layer separating the nanowire from the reflector is $243\,\mathrm{nm}$ thick, which forms a stripline structure with the reflector acting as a ground plane. In Fig.\,\ref{supfig:microwave} we show the simulation of the characteristic impedance and phase velocity factor for the full nanowire stripline. The geometrical parameters are shown in the cross-sectional sketch. By assuming a niobium nitride sheet kinetic inductance $L_\mathrm{k}=80\,\mathrm{pH}$ per square (typical of $7\,\mathrm{nm}$ thick layer), a $100\,\mathrm{nm}$-wide wire has characteristic impedance $Z_0=3261\,\mathrm{\Omega}$ and a phase velocity $1.36\,\%$ of the speed of light. 

As explained in Sec.\,\ref{sec_SNSPD_design}, in order to preserve the signal-to-noise ratio and the timing information of the output pulses, microwave impedance matching is required. We designed and monolithically integrated two Hecken tapers, one for each side of the differential nanowire (Fig.\,\ref{fig:architecture}(a)(i)). The tapers were designed with Phidl CAD layout software\,\cite{phidl}. To reduce the length of the taper and the fabrication complexity we started the matching from a width of $300\,\mathrm{nm}$. The $300\,\mathrm{nm}$ and $100\,\mathrm{nm}$ wire sections were connected through optimal curves (Fig.\,\ref{fig:architecture}(a)(ii)). The tapers have a total length of $7.6\,\mathrm{mm}$ with $8648$ squares, a kinetic inductance of $0.69\,\mathrm{\mu H}$, a $537\,\mathrm{MHz}$ lower cutoff frequency, and it introduces a $0.93\,\mathrm{ns}$ delay. Note that for the MIT devices, the layout was patterned with a positive tone resist, therefore resembling a grounded coplanar waveguide (CPW) geometry after etching. Because the width of the slot of the CPW is one order of magnitude larger than the thickness of the silicon oxide layer, the structure behaves approximately as a stripline as the side grounds can be neglected.

\section{Spice simulation}
\label{app_spice}
\begin{figure}[!ht]
    \centering
    \includegraphics[width=\linewidth]{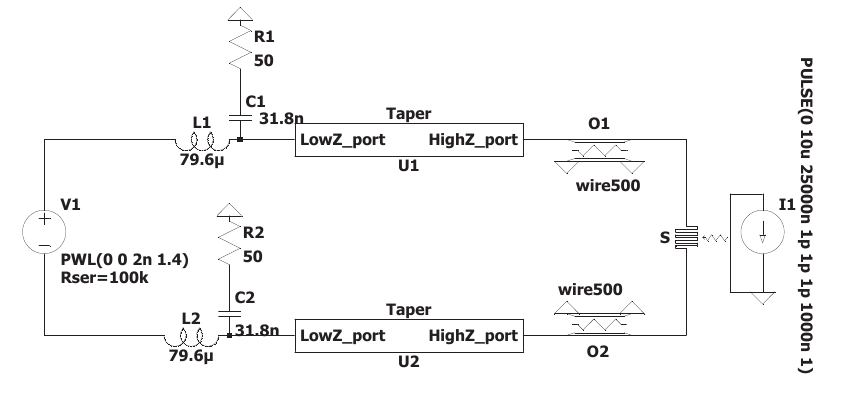}
    \caption{Schematics of the SPICE simulation for an impedance matched differential detector. The taper is modeled as cascaded transmission lines (300 sections) with varying impedance and phase velocities. The nanowire is simulated as a lossy transmission line (LTRA) with $L=800.6\,\mathrm{\mu H/m}$ and $C=75.27\,\mathrm{pF/m}$. In this specific scheme the length of the symmetric transmission line are set to $500\,\mathrm{\mu m}$. Note that L1, C1, L2, and C2 constitute bias-tee elements. Their values were selected arbitrarily. }
    \label{supfig:spice}
\end{figure}

We simulated the impedance matched differential detector using a SPICE model that incorporates both the electrothermal feedback of the nanowire and microwave dynamics of the line. The SPICE model of the SNSPDs
is implemented by Berggren et \textit{al.} \cite{berggren2018superconducting} based on the phenomenological hotspot velocity model
by Kerman et \textit{al.} \cite{kerman2009electrothermal}. The nanowire transmission line is simulated using a SPICE lossy transmission line model (LTRA) with capacitance per unit length and inductance per unit length values reproducing the impedance and phase velocity of the line ($L=800.6\,\mathrm{\mu H/m}$ and $C=75.27\,\mathrm{pF/m}$). We did not include any loss term. The taper was modeled as a cascade (300 sections) of transmission line sections with varying impedance and phase velocities. To simulate photon detection in different sections of the meander line we changed the length of the nanowire transmission line. In the example of Fig. \ref{supfig:spice}, we assume detection at the center of the meander (the two nanowire transmission line are symmetric).

\section{Nanofabrication} \label{fab}

We fabricated our differential SNSPDs on $100\,\mathrm{mm}$ silicon wafers, using a Canon EX3 DUV stepper \footnotemark[1] to fabricate around 200 SNSPDs per wafer. We began by lifting off a patterned Ti/Au/Ti mirror, with $80\,\mathrm{mm}$ thick Au. The Ti layers improved adhesion between the metals and dielectrics but were kept thin ($\approx 1\,\mathrm{nm}$ to $2\,\mathrm{nm}$) due to the better performance of Au as a mirror around $1550\,\mathrm{nm}$. A spacer layer of SiO$_2$ was then blanket RF-bias sputtered on top of the mirrors while biasing the substrate holder at $375\,\mathrm{kHz}$ to produce a smoother SiO$_2$ film. The JPL NbN superconducting layer was subsequently deposited by reactive-sputtering a Nb sputtering target in a N$_{2}$/Ar gas mixture, while applying RF-bias to the substrate to reduce the grain size in the films following Ref.\cite{dane2017bias}. The DC power applied to the Nb target was $340\,\mathrm{W}$, and the RF-bias was $6\,\mathrm{W}$ to $7\,\mathrm{W}$. The films were deposited over a fixed time of 13 seconds, which gave an approximately $7\,\mathrm{nm}$ thick film based on ellipsometry of similar films on SiO$_2$-on-Si wafers. NbN films at MIT were made using the same setup as in Ref.\,\cite{dane2017bias}. The resulting film was approximately $7\,\mathrm{nm}$, determined through measurements of the sheet resistance and compared to similar films deposited on SiO$_2$/Si. We next fabricated Ti/Au/Ti bond pads, using ion milling prior to Ti/Au/Ti e-beam evaporation in order to produce good contact to the NbN. We employed electron-beam lithography to write the nanowires and tapers simultaneously. The wafer with JPL NbN used negative-tone resist while the MIT wafer used positive-tone resist. The JPL NbN films were etched in a mixture of CCl$_2$F$_2$/CF$_4$/O$_2$ in an inductively coupled plasma reactive ion etcher (ICP-RIE), while MIT etched using CF$_{4}$ plasma. A blanket film of $\approx 100\,\mathrm{nm}$ of SiO$_2$ was deposited to protect the SNSPDs immediately after this etch and the removal of the e-beam photoresist in solvent baths. We then exposed a liftoff pattern in the stepper to define the remainder of the anti-reflection (AR) stack above the active area of the SNSPDs but away from the bond pads, to avoid wire bonding through thick dielectrics. The AR stacks on each of the three wafers were simulated using refractive indices data of each layer and designed to maximize efficiency at $1550\,\mathrm{nm}$.  For two of the wafers, we fabricated a one-layer AR stack of SiO$_2$/TiO$_2$.  For the third wafer considered here, we used a double-layer AR stack of SiO$_2$/TiO$_2$ to produce a narrower band around $1550\,\mathrm{nm}$. The one-layer AR stack was $271\,\mathrm{nm}$/$167\,\mathrm{nm}$  (SiO$_2$/TiO$_2$) for one wafer with the JPL NbN film. Meanwhile, a wafer with MIT NbN had a stack of $263\,\mathrm{nm}$/$184\,\mathrm{nm}$. The thicknesses differed due to the difference in refractive indices between the JPL and MIT NbN films. The two-layer AR stack on the other wafer with JPL NbN was approximately $150 \,\mathrm{nm}$/$279\,\mathrm{nm}$/$157 \,\mathrm{nm}$/$262 \,\mathrm{nm}$ (SiO$_2$/TiO$_2$/SiO$_2$/TiO$_2$).  Next, we wrote an etch-back pattern with the stepper and used the ICP-RIE with CHF$_3$ and O$_2$ to etch through the blanket SiO$_2$ and spacer SiO$_2$. Finally, we exposed an etch-back pattern and used deep reactive ion etching (DRIE) to define the lollipop pattern that allowed the detectors to be released from the wafer and inserted into zirconia sleeves for self-alignment to single-mode optical fibers \cite{Verma_IEEE_2012}. We screened each detector at room temperature by probing the electrical resistance and measuring the reflection spectrum to determine the characteristics of the full optical cavity encompassing the mirror, spacer layer, NbN nanowires, and AR stack, before testing.

\section{Cryogenic readout} \label{cryoamp}

Thanks to the impedance matched design of the detectors, the signal amplitude is significantly increased, by as much as a factor of three compared to a regular SNSPD\,\cite{zhu2019superconducting}. This results in a signal amplitude in the range of a few millivolts for both positive and negative pulses, at the input of the first amplifier. To prevent saturation of the amplifier, a high dynamic range, single-stage cryogenic amplifier was developed by Cosmic Microwave Technology \footnotemark[1]. The amplifier achieves a gain of 24\,dB and a bandwidth of 2\,GHz, while exhibiting a noise temperature of 4\,K to 7\,K between 1\,MHz and 2\,GHz. The input power for 1\,dB compression is $-24\,\mathrm{dBm}$. By using two of these amplifiers, one of each end of the differential detector (see Appendix \ref{app_setup} for details on the measurement setup), we verified that the dynamic range was sufficient to prevent saturation of the detector signal and that it maintained uniform gain for both positive and negative signals (see Fig.\,\ref{fig:impedance_matched_signal}).

To achieve cancellation of the geometric jitter with a differential comparator (see Fig.\,\ref{fig:triggering}) a SiGe current-mode logic comparator was used at the 45\,K stage of the cryostat and the offset of the input pulses was achieved with a pair of bias-tees between the cryogenic amplifiers and the comparator (see Appendix \ref{app_setup} for details on the measurement setup)

\section{Measurement Setup} \label{app_setup}
\begin{figure*}
    \centering
    \includegraphics[width=\linewidth]{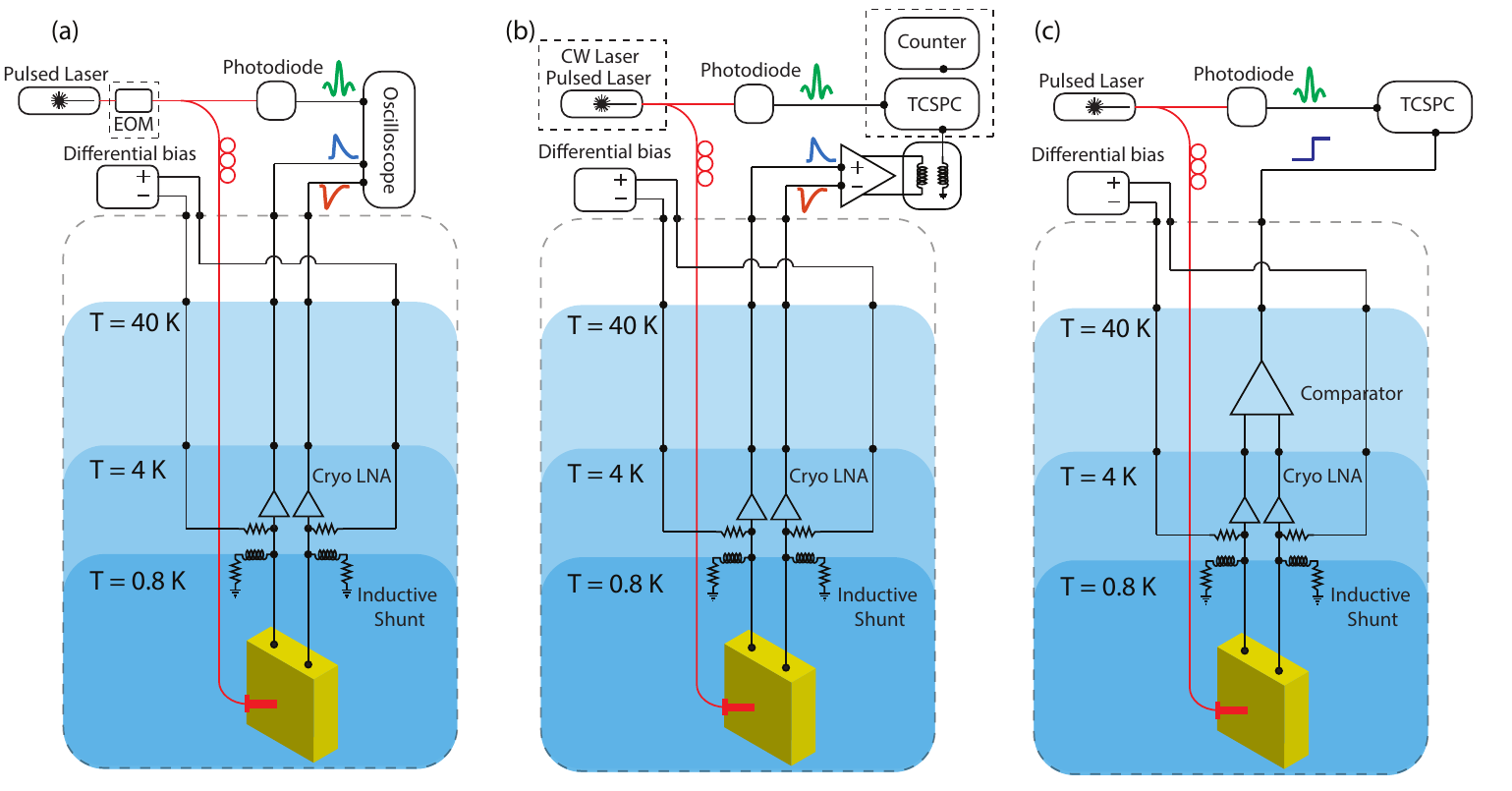}
    \caption{Measurement setups. (a) Measurement setup for the characterization of the detector pulses, detector jitter $t_\Sigma$, and differential time $t_\mathrm{\Delta}$-histogram. (b) Measurement setup for the characterization of the system detection efficiency and for the measurement of the system jitter $j_\mathrm{diff}$ using the balun in combination with the TCSPC module. (c) Measurement setup for the characterization of the system jitter $j_\mathrm{diff}$ using the cryogenic comparator in combination with the TCSPC module.}
    \label{supfig:setup}
\end{figure*}

\begin{table*}
	\centering
	\begin{tabular}{c|c}
		Reference & Instrument\footnotemark[1] \\ \hline \hline
		Cryogenic LNA & CMT LF1S ($1\,\si{MHz} - 2\,\si{GHz}$) \\ \hline
		Real-time oscilloscope & Keysight DSOZ634A ($63\,\mathrm{GHz}$) \\ \hline		
	    Differential Amplifier & Analog Devices LTC6432-15 ($100\,\si{kHz} - 1.4\,\si{GHz}$) \\ \hline
		Balun board & Texas Instruments ADC-WB-BB/NOPB ($4.5\,\si{MHz} - 3\,\si{GHz}$) \\  \hline
	    Cryogenic Comparator & Analog Devices HMC675LP3E  \\  \hline
	    Inductive shunts & Custom: $1.1\,\mathrm{\mu H}$ + $50\,\mathrm{\Omega}$  \\  \hline
	    TCSPC module & Becker \& Hickl SPC-150NXX  \\  \hline
	    Photodiode & New Focus 1014  (45 GHz) \\  \hline
	    Pulsed Laser & Calmar Mendocino $1550\,\mathrm{nm}$ $10\,\mathrm{MHz}$ repetition rate  \\  \hline
	    Universal Counter & Keysight 53220A \\  \hline
	    Amplitude Modulator & iXblue MXER-LN-20 \\  \hline
	\end{tabular}
	\caption{Overview of the instruments}
	\label{tab:inst_overview}
\end{table*}
Figure \ref{supfig:setup} shows the experimental setups used for the characterization of our differential detectors. 
Table \ref{tab:inst_overview} provides an overview of the instrumentation\,\footnotemark[1]. In all the setups, the detectors were biased with a fully differential circuit. Moreover, to bias as close as possible to switching current and avoid latching at high photon fluxes we added cryogenic inductive shunts at both ports. 

Figure \ref{supfig:setup}(a) shows the experimental setup used for the characterization of the detector pulses, detector jitter $t_\Sigma$, differential time $t_\mathrm{\Delta}$-histogram, and photon number resolution capabilities. After amplification with two low-noise cryogenic amplifiers (CMT LF1S)\footnotemark[1], the detector outputs are directly interfaced with a real-time oscilloscope (Keysight DSOZ634A)\footnotemark[1], which we used to acquire the traces for post-processing. To improve the SNR and reduce the impact of the electrical noise the trigger was set on the steepest part of the pulse rising edge. The oscilloscope sampling rate was set to $80 \times 10^9 \,\mathrm{s^{-1}}$ to minimize quantization error. The analog bandwidth was set to $6\,\si{GHz}$\,\cite{korzh2020demonstration}. For measuring the PNR capabilities, the pulsed laser repetition rate was reduced to $1\,\mathrm{MHz}$ using an intensity modulator.

The setup shown in Fig \ref{supfig:setup}(b) was used for the characterization of the system detection efficiency and for the measurement of the system jitter $j_\mathrm{diff}$ using the balun in combination with the TCSPC module. After amplification with the cryogenic low noise amplifier, the detector outputs undergo a further stage of amplification through a low noise differential amplifier (LTC6432-15)\footnotemark[1]. This is necessary to improve the signal level before the balun, which has a $6\,\mathrm{dB}$ insertion loss. From the differential amplifier the outputs are connected to the balun board (ADC-WB-BB)\footnotemark[1], which performs the difference of the complementary pulses. For the characterization of the detection efficiency, the output of the balun was connected to a universal counter (Keysight 53220A)\footnotemark[1]. The detection efficiency was characterized after a calibration of the optical path losses. To measure the system jitter $j_\mathrm{diff}$, the output of the balun was connected to the TCSPC module (B\&H SPC-150NXX) \footnotemark[1] together with a synchronization signal from the pulsed laser (Calmar Mendocino, $1550\,\si{nm}$)\footnotemark[1] obtained with a fast photodiode.

The setup shown in Fig \ref{supfig:setup}(c) was used for the characterization of the system jitter $j_\mathrm{diff}$ using the cryogenic comparator in combination with the TCSPC module. After amplification with the cryogenic low noise amplifier, the detector outputs are fed to the differential comparator. Its output is connected to the TCSPC module together with the synchronization signal.

\section{Additional details on differential delay-line imaging capabilities} \label{form_dev}

\begin{figure}[!ht]
    \centering
    \includegraphics{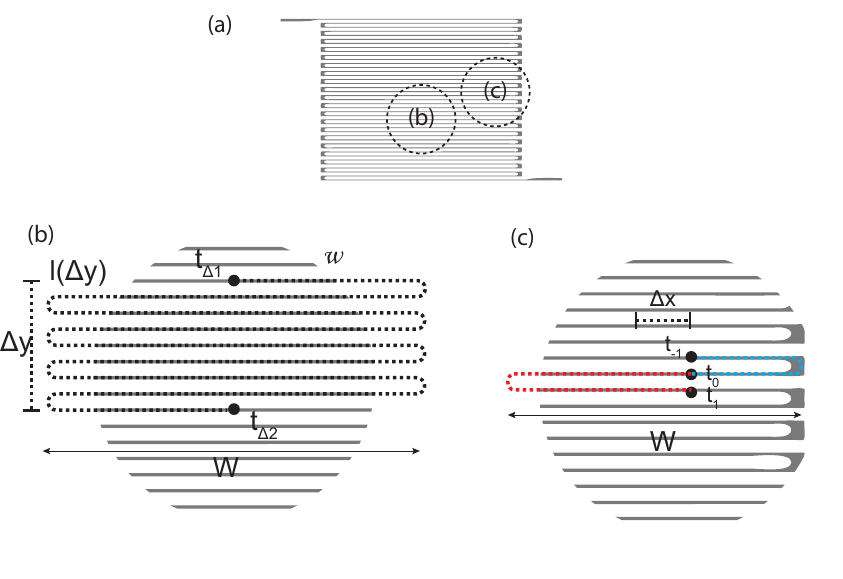}
    \caption{Sketches for derivation of delay-line imaging formulas. (a) Fiber spot misalignment reference on the meander for (b) and (c). (b) The spot is vertically misaligned. (c) The spot is horizontally misaligned.}
    \label{fig_form_derivation}
\end{figure}

In this Section we derive the first order formula introduced in the main text to connect the characteristics of the $t_\Delta$-histogram with effective mode misalignment on the meander $(\Delta x,\Delta y)$.
In Fig. \ref{fig_form_derivation}(b) we analyze the condition in which the mode is vertically misaligned.
In general
\begin{align}
\begin{split}
t_{\Delta,1}&=\frac{L_\mathrm{m}-2x_\mathrm{p}}{v_{\mathrm{ph}}} \\
t_{\Delta,2}&=\frac{L_\mathrm{m}-2(x_\mathrm{p}+l(\Delta y))}{v_{\mathrm{ph}}} \\
\end{split}
\end{align}
where $l(\Delta y)=W \frac{w \Delta y}{\mathrm{FF}}$ is the linearized length on the meader corresponsing to $\Delta y$, as shown in the Fig.\ref{fig_form_derivation}(a). Here $w$ the width of the nanowire, $W$ the width of the meander, and $\mathrm{FF}$ the meander fill-factor.
The time-domain shift associated to $l(\Delta y)$ is 
\begin{align}\label{deltay}
t_\mathrm{y}=t_{\Delta,2}-t_{\Delta,1}=\frac{2l(\Delta y)}{v_{\mathrm{ph}}}=\frac{2W}{v_{\mathrm{ph}}}\frac{w \Delta y}{\mathrm{FF}}
\end{align}   

When $\Delta y =0$, the peak of the envelope of the $t_\Delta$-histogram is at $t_{\Delta,1}=0$. By rearranging Eq. \ref{deltay} we obtain a first order formula to evaluate the vertical shift in function of the peak of the envelope of the $t_\Delta$-distribution.
Note that this derivation assumes $\Delta x = 0$, and it's therefore accurate within one meander pitch.
In Fig. \ref{fig_form_derivation}(c) we analyze the situation in which the mode is horizontally misaligned.In general
\begin{align}
\begin{split}
t_{\Delta,0}&=\frac{L_\mathrm{m}-2x_\mathrm{p}}{v_{\mathrm{ph}}} \\
t_{\Delta,-1}&=\frac{L_\mathrm{m}-2\left[x_\mathrm{p}-(W-2\Delta x)\right]}{v_{\mathrm{ph}}}
\\
t_{\Delta,1}&=\frac{L_\mathrm{m}-2\left[x_\mathrm{p}+(W+2\Delta x)\right]}{v_{\mathrm{ph}}} \\
\end{split}
\end{align}
The relative difference between adjacent histogram sub-peaks is:
\begin{align}
\begin{split}\label{deltax}
t_x &= \Delta t_{\Delta, 1,0} - \Delta t_{\Delta, 0,-1} \\
&= (t_{\Delta,1}-t_{\Delta,0} )-(t_{\Delta,0}-t_{\Delta,-1})\\
&=t_{\Delta,1} - 2t_{\Delta,0} + t_{\Delta,-1} \\
& = \frac{8\Delta x}{v_{\mathrm{ph}}}
\end{split}
\end{align}
By rearranging Eq. \ref{deltax} we obtain a first order formula to evaluate the horizontal shift in function of the relative difference between adjacent histogram sub-peaks.

\section{Additional details on Photon Number Resolution} \label{app_pnr}
\begin{figure}[!ht]
    \centering
    \includegraphics{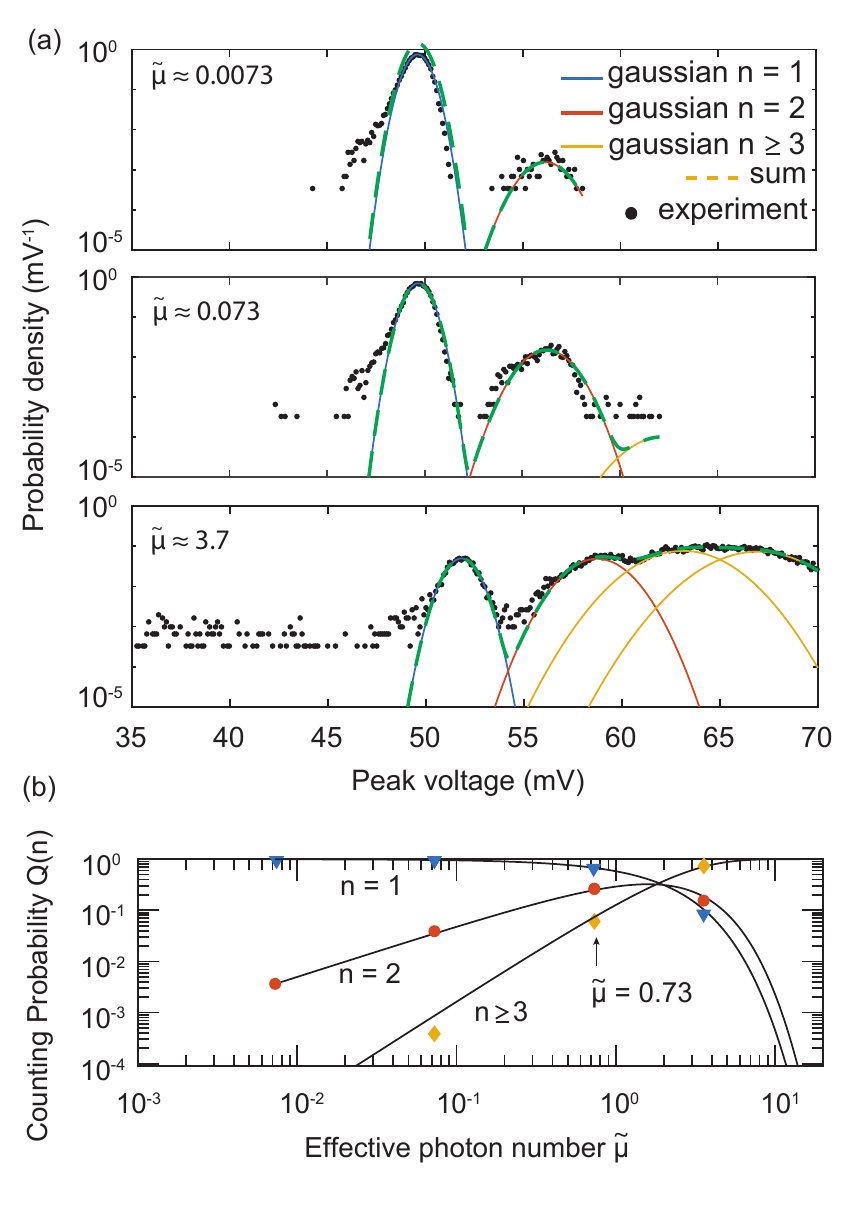}
    \caption{Photon number resolution. (a) Gaussian fitting of the pulse amplitude histograms for several effective photon numbers. (B) Photon counting statistics reconstructed from the pulse height distributions under different illumination conditions.}
    \label{supfig:pnrfit}
\end{figure}

We characterized the photon number resolution capabilities of our detector. We used the setup shown in Fig. \ref{supfig:setup}(a). The repetition rate of our $1550\,\mathrm{nm}$ pulsed laser was $10\,\mathrm{MHz}$. The reset time of our detector was $\approx 160\,\mathrm{ns}$. We used an intensity modulator to reduce the repetition rate of the laser to $1\,\mathrm{MHz}$ to allow the detector to fully reset after a detection event. We estimated the effective photon number by measuring the output photon rate and scaling by the system detection efficiency and the repetition rate. 
\begin{equation}
\tilde{\mu}=\frac{\text{photon rate}}{\text{rep. rate}}\,\,\mathrm{SDE}
\end{equation}

Fig. \ref{supfig:pnrfit}(a) shows the distribution of the peak of the difference of the complementary pulses for several effective photon numbers. Each distribution was fitted with up to four Gaussian functions. The counting statistics are reconstructed by integrating the areas under each Gaussian. Notice that the fourth Gaussian was introduced to fit the distribution for $\tilde{\mu}=3.7$. We grouped the counting probability for $n\geq 3$ as these events were not well-separated. For every effective mean photon number, the left shoulder was excluded from the fit. Note that for increasing photon number the amplitude of the pulses increases. 

In Fig.\,\ref{supfig:pnrfit}(b) we plot the counting probability $Q(n)$ (markers), extracted by integrating the area under each Gaussian distribution, with the photon statistics of the coherent source $S(n) =e^{-\tilde{\mu}}\tilde{\mu}^n/n!$ (line). We grouped the probability for the events with $n \geq 3$, which are not clearly separated. Note that in the figure we normalized the theoretical $S(n)$ by the probability of zero photons $S(n)/(1-S(0))$ where $S(0)=e^{-\tilde{\mu}}$. Further details on the procedure are available in Ref.\,\cite{zhu2020resolving}.

\section{Estimation of uncertainties}
\label{app_uncertainty}
The uncertainties on the detector jitter $j_\Sigma$ are estimated as the $95\,\%$ confidence bound on the data. The uncertainties on the system jitter $j_\mathrm{diff}$ are mainly due to the resolution of the TCSPC module and estimated as two time-bins $\sigma_\mathrm{TCSPC}=0.4\,\mathrm{ps}$. 

The system detection efficiency is estimated as
\begin{equation}
    \text{SDE}=\frac{\text{count rate}}{\text{photon flux}}.
\end{equation}
We assume the uncertainty on the count rate to be negligible. The photon flux is estimated as:
\begin{equation}
    \text{photon flux}=\frac{P_0 A_1 A_2 A_3}{E_\lambda}
\end{equation}
where $P_0$ is the input optical power, measured with a calibrated power meter, $A_{1-3}$ are the values of three optical attenuators used to attenuate
the optical power, and $E_\lambda$ is the energy of the photon of wavelength $\lambda$.
To measure the attenuation ratios $A_{1-3}$, the attenuators are connected in series and interfaced to the calibrated power meter through an optical switch. To measure the  attenuation ratio we set one of the attenuator to the desired values and the others to $0\,\mathrm{dB}$. We measure the attenuated output power and we repeat the procedure for the other attenuators. We did not calibrate the optical switch. $A_{1-3}$ result from a relative measurement of the optical power using the same power meter. Therefore, to estimate the uncertainty on $A_{1-3}$ we only consider the relative uncertainty due to the non-linearity $\sigma_{\mathrm{NL}}(P)/P=0.5\%$, on each measurement.
The total uncertainty on the SDE is dominated by the power meter uncertainty $\sigma(P)/P=5\%$, and it is approximately $\sigma(\mathrm{SDE})/\mathrm{SDE}=5.2\%$.
\bibliography{biblio_diff_single_up}

\end{document}